\RequirePackage{fix-cm}
\documentclass[smallextended]{svjour3}       
\usepackage{cite}
\usepackage{amsmath,amssymb,amsfonts}
\usepackage{algorithmic}
\usepackage{graphicx}
\usepackage{textcomp}
\usepackage{xcolor}
\usepackage{nicefrac}
\usepackage{graphics}
\usepackage{float}
\usepackage{graphicx}
\usepackage{amsmath}
\usepackage{color}
\usepackage{dsfont}
\usepackage{bm}
\usepackage{upgreek}
\usepackage{arydshln}
\usepackage{enumerate}
\usepackage{amsthm}
\usepackage{graphicx}
\usepackage{threeparttable}
\usepackage{multirow}
\usepackage{color}
\usepackage{xfrac}
\usepackage{array}
\usepackage{hyphenat}
\usepackage{algorithmic}
\usepackage{graphicx}
\usepackage{textcomp}
\usepackage{xcolor}
\usepackage{algorithm,algorithmic}
\usepackage[numbers]{natbib}
\usepackage{url}
\newcommand{\be}{\begin{equation}}
\newcommand{\ee}{\end{equation}}

\newcommand{\norm}[1]{ || #1 ||}
\newcommand{\trace}[1]{ \mathrm{tr}( #1 )}
\newcommand{\mb}[1]{\mathbf{#1}}
\newcommand{\bs}[1]{\boldsymbol{#1}}

\newcommand{\virg}[1]{\textquotedblleft#1\textquotedblright}

\newcommand{\kronVtinvmT}{\mb{V}_1^{-\top/2}\otimes\mb{V}_1^{-1/2}}

\newcommand{\topH}{\mathsf{H}}

\newcommand{\pV}{\widehat{\mb{V}}^\star_1}

\newcommand{\vc}[1]{\mathrm{vec}( #1 )}

\newcommand{\cvec}[1]{ \mathrm{vec}\left(  #1 \right) }

\newcommand{\ovec}[1]{\underline{\mathrm{vec}}(#1)}

\newcommand{\tonde}[1]{\left( #1 \right)  }
\newcommand{\quadre}[1]{\left[  #1 \right]  }
\newcommand{\graffe}[1]{\left\lbrace   #1 \right\rbrace   }

\newcommand{\kronVtinvmTS}{(\pV)^{-\top/2} \otimes (\pV)^{-1/2}}
\newcommand{\kronVtinvTS}{(\pV)^{-\top} \otimes (\pV)^{-1}}
\newcommand{\kronVTS}{(\pV)^{\top} \otimes \pV}

\journalname{Journal of Signal Processing Systems}
\begin{document}

\title{Joint Estimation of Location and Scatter in Complex Elliptical Distributions\thanks{The work of S. Fortunati, A. Renaux and F. Pascal has been partially supported by DGA under grant ANR-17-ASTR-0015.}
}
\subtitle{A robust semiparametric and computationally efficient $R$-estimator of the shape matrix}


\author{Stefano~Fortunati         \and
        Alexandre~Renaux  \and
        Fr\'{e}d\'{e}ric~Pascal
}


\institute{Stefano~Fortunati \at
              Universit\'{e} Paris-Saclay, CNRS, CentraleSup\'{e}lec, Laboratoire des signaux et syst\`{e}mes, 91190, Gif-sur-Yvette, France \&
              DR2I-IPSA, 94200, Ivry sur Seine, France\\
              \email{stefano.fortunati@centralesupelec.fr}           
           \and
           Alexandre Renaux and Fr\'{e}d\'{e}ric~Pascal \at
           Universit\'{e} Paris-Saclay, CNRS, CentraleSup\'{e}lec, Laboratoire des signaux et syst\`{e}mes, 91190, Gif-sur-Yvette, France\\
           \email{alexandre.renaux@universite-paris-saclay.fr, frederic.pascal@centralesupelec.fr}.
}

\date{Received: date / Accepted: date}

\maketitle

\begin{abstract}
The joint estimation of the location vector and the shape matrix of a set of independent and identically Complex Elliptically Symmetric (CES) distributed observations is investigated from both the theoretical and computational viewpoints. This joint estimation problem is framed in the original context of semiparametric models allowing us to handle the (generally unknown) density generator as an \textit{infinite-dimensional} nuisance parameter. In the first part of the paper, a computationally efficient and memory saving implementation of the robust and semiparmaetric efficient $R$-estimator for shape matrices is derived. Building upon this result, in the second part, a joint estimator, relying on the Tyler's $M$-estimator of location and on the $R$-estimator of shape matrix, is proposed and its Mean Squared Error (MSE) performance compared with the Semiparametric Cram\'{e}r-Rao Bound (CSCRB).
\end{abstract}

\section{Introduction}
\label{sec:intro}

Inferring the correlation structure of a set of centered data if a key step in many signal processing and machine learning procedures. Among others, radar/sonar detection, image segmentation, dimension reduction, distance learning and clustering rely on the estimation of the covariance/correlation matrix of an acquired data set \cite{bishop2007}. Along with the need of an estimation of the covariance matrix, there is another common aspect in all the above-mentioned applications: the non-Gaussian and heavy-tailed nature of the data. As a consequence, popular Gaussian and pseudo-Gaussian inference procedures may present a dramatic performance decay as extensively shown in statistics and signal processing literature (see e.g. \cite{book_zoubir} and the references therein).

Motivated by a wide range of experimental evidences and measurement campaigns, the (Real or Complex) Elliptically Symmetric (ES) model has been recently adopted to characterize the statistical data behavior. The RES and CES models in fact have been proved to be able to catch the data heavy-tailedness in a large variety of applications such us radar/sonar \cite{Esa_clutter,Sang}, hyper-spectral imaging \cite{Joana,Mano} and clustering \cite{violeta,Muma_arx} just to cite a few. From now on, in this paper, we will focus our attention only on complex-valued,  CES distributed, datasets. This choice allows us to work in the most general framework, since the obtained results can be easily \virg{brought back} to the real-valued case.     

Together with its generality, the second feature that has placed the elliptical model in the spotlight of signal processing and machine learning communities is its \virg{parsimony} in terms of required parameters. In fact, CES model is fully specified by two \textit{finite dimensional} parameters, i.e. a location vector and a covariance/scatter matrix (as the classical Gaussian model) and by an \textit{infinite dimensional} functional parameter, usually called \textit{density generator}, characterizing the data heavy-tailedness. To better understand the role of this infinite dimensional term, let us take a step back to introduce the notion of \textit{semiparametric} model. 

Let $\{\mb{z}_l\}_{l=1}^L$ be a set of $L$ independent and identically distributed (i.i.d.)\ $N$-dimensional observations sharing the same probability density function (pdf), i.e.\ $\mathbb{C}^N \ni \mb{z}_l \sim p_Z,\; \forall l$. A parametric model $\mathcal{P}_{\bs{\theta}} \triangleq \graffe{p_Z\left| p_Z(\mb{z}_l;\bs{\theta}), \bs{\theta} \in \Theta\right. }$ is then defined as a family of pdfs parameterized by a finite dimensional vector $\bs{\theta} \in \Theta$. As a classical example, in multivariate Gaussian-based inference, $\bs{\theta}$ is set up by the mean vector $\bs{\mu}$ and by the covariance/scatter matrix $\bs{\Sigma}$. 

However, a parametric model is generally too \virg{narrow} and it fails to take into account all the actual uncertainty about the data distribution that is generally present in practical scenarios. Semiparametric models have been then introduced to provide with an additional (functional) degree of freedom \cite{BKRW}. A semiparametric model $\mathcal{P}_{\bs{\theta},h} \triangleq \graffe{p_Z\left| p_Z(\mb{z}_l;\bs{\theta},h), \bs{\theta} \in \Theta, h\in \mathcal{G} \right. }$ is a family of pdf parameterized by a finite dimensional vector $\bs{\theta} \in \Theta$ (as in the classical parametric case) and by a function $h$ belonging to some suitable function space $\mathcal{G}$. Usually, in most applications, $\bs{\theta} \in \Theta$ is the parameter vector of interest, while $ h\in \mathcal{G}$ can be considered as a \textit{nuisance} function that \virg{contains} the missing knowledge of the functional form of the data pdf $p_{Z}$. Consequently, inference procedures in semiparametric models aim at estimating/testing for $\bs{\theta} \in \Theta$ in the presence of an unknown function $ h\in \mathcal{G}$ whose estimation is not strictly required.       

It is now immediate to realize that the CES model can be framed as a semiparametric model \cite{For_SCRB_complex,Hallin_P_Annals,Hallin_Annals_Stat_2}. Formally, the pdf $p_Z$ of a CES-distributed random vector $\mb{z}_l \in \mathbb{C}^N$ can be expressed as \cite{Esa}:
\be\label{CES_pdf}
p_Z(\mb{z}_l|\bs{\mu},\bs{\Sigma},h)=|\bs{\Sigma}|^{-1} h \left((\mb{z}_l-\bs{\mu})^\mathsf{H}\bs{\Sigma}^{-1}(\mb{z}_l-\bs{\mu}) \right),
\ee
where, as said before, the final dimensional parameter of interest $\bs{\theta} \in \Theta$ is composed of a location vector $\bs{\mu}$ and by the scatter matrix $\bs{\Sigma}$ that represents the correlation structure of the data, while the \textit{nuisance} density generator $h$ belongs to the set
\be\label{set_G}
\mathcal{G} = \graffe{ h:\mathbb{R}^+\rightarrow \mathbb{R}^+ \left| \int_{0}^{\infty}t^{N-1}h(t)dt < \infty, \int p_Z =1 \right. }.
\ee

Two considerations are now in order:
\begin{itemize}
	\item \textit{The identifiability issue}: the scatter matrix $\bs{\Sigma}$ and the density generator $h$ are not jointly identifiable. Consequently, only scaled versions, usually called \textit{shape matrix}, $\mb{V}\triangleq \bs{\Sigma}/s(\bs{\Sigma})$ can be estimated \cite{Esa}. According to our recent work \cite{Sem_eff_est_TSP}, from now on we consider the shape matrix 
	\be\label{const_11}
	\mb{V}_1\triangleq \bs{\Sigma}/[\bs{\Sigma}]_{1,1},
	\ee
	i.e.\ the one obtained form the scatter matrix by constraining its first top-left element to be equal to one.
	\item \textit{Augmented complex representation of $\bs{\theta}$}: Following the rules of the Wirtinger calculus \cite{Complex_M,Remmert,Kreutz}, in order to take into account the complex-value nature of the location vector $\bs{\mu}$ and of the shape matrix $\mb{V}_1$, the finite-dimensional parameter $\bs{\theta}$ has to be built up as \cite{Sem_eff_est_TSP}: \footnote{The operator $\ovec{\mb{A}}$ defines the $N^2-1$-dimensional vector obtained from $\cvec{\mb{A}}$ by deleting its first element, i.e.\ $\cvec{\mb{A}} \triangleq [a_{11},\ovec{\mb{A}}^T]^T$. In general, in this paper, we always adopt the same notation used in our previous work \cite{Sem_eff_est_TSP}.}
	\be
	\label{theta_com}
	\bs{\theta} \triangleq (\bs{\mu}^\top,\bs{\mu}^\topH,\ovec{\mb{V}_1}^\top)^\top \in \Theta \subseteq \mathcal{C}^q,
	\ee       
	where $q=N(N+2)-1\;(=2N+N^2-1)$ and $\mathcal{C}^q$ is a complex-vector space on real field of dimension $q$ \cite{Complex_M,Kreutz}. Note that the \virg{$-1$} is due to the fact that we constraint the first top-left element of $\mb{V}_1$ to be equal to 1, so it does not have to be estimated.
\end{itemize}

Building upon the previous considerations, the semiparamentric CES model \cite{For_SCRB_complex} can be cast as:
\be
\label{CES_semi_par_model}
\mathcal{P}_{\bs{\theta},h} = \left\lbrace  p_Z | p_Z(\mb{z}_l|\bs{\theta},h) = |\mb{V}_1|^{-1} h \left((\mb{z}_l-\bs{\mu})^\topH\mb{V}_1^{-1}(\mb{z}_l-\bs{\mu}) \right);  \bs{\theta} \in \Theta, h \in \mathcal{G} \right\rbrace,
\ee 

Estimating $\bs{\theta} \in \Theta$ in the presence of different \virg{degrees of uncertainty} on the density generator $h$ is a well-known problem in robust statistics and signal processing. The most popular class of robust estimators for $\bs{\theta} \in \Theta$ belongs to the family of $M$-estimators \cite{huber_book} and has been firstly proposed by Maronna in \cite{maronna1976} and further developed and investigated by Tyler \cite{Tyler1}. However, if on one hand the Maronna/Tyler $M$-estimators have the remarkable robustness property, they fail to be \textit{semiparametrically efficient}, i.e their Mean Square Error (MSE) does not achieve the Semiparametric Cram\'{e}r-Rao Bound \cite{For_SCRB,For_SCRB_complex}.

In order to fill this gap, in their seminal work Hallin, Oja and Paindaveine \cite{Hallin_Annals_Stat_2} proposed a new class of rank-based $R$-estimators of the shape matrix for a set of centered RES-distributed data able to be both \textit{distributionally robust} and (almost) \textit{semiparametric efficient}. \footnote{The interested reader can find the Matlab and Python code related to our implementation of this $R$-estimator in both RES- and CES- distributed data at \url{https://github.com/StefanoFor}.} The real-valued $R$-estimator proposed in \cite{Hallin_Annals_Stat_2} has been expended to the case of CES-distributed data in our recent work \cite{Sem_eff_est_TSP} where a theoretical and simulative analysis of its \virg{finite-sample} performance has been provided as well.

Following the trail of \cite{Hallin_Annals_Stat_2,Sem_eff_est_TSP}, the present paper has two main goals:
\begin{enumerate}
	\item Derive a \virg{computationally efficient} version of the complex-valued shape matrix $R$-estimator proposed in \cite{Sem_eff_est_TSP},
	\item Investigate the joint estimation problem of the location parameter $\bs{\mu}$ and the shape matrix $\mb{V}_1$ in the presence of an unknown density generator $h$. \footnote{This part has been partially addressed in our related conference paper \cite{For_MLSP}.}
\end{enumerate}

Having a \virg{computationally efficient} implementation of an estimator is of fundamental importance in real-time applications or in high-dimensional data sets. The new version of the $R$-estimator of the shape matrix proposed in Section \ref{sec_comp_eff_R} of this paper is computationally faster and \virg{memory saving} than the previous implementation in \cite{Sem_eff_est_TSP} making its exploitation possible in a wider range of applications. The paper continues with an exhaustive theoretical investigation of the statistical interrelation underlying the joint semiparametric estimation of the location vector $\bs{\mu}$ and of the shape matrix $\mb{V}_1$ provided in Section \ref{joint_est} while a robust semiparametric (and computationally) efficient joint estimator of location and shape is discussed in Section \ref{rob_sem_eff_est}. The numerical analysis of the performance of the proposed joint estimator is presented in Section \ref{sec:num}. Finally, some concluding remarks are collected in Section \ref{conclusions}.

We conclude this nonproductive Section with three paragraphs summarizing some useful notations and definitions that serve as prerequisites to the comprehension of the material presented in the rest of the paper.
  
\textit{Algebraic notation}: For the sake of consistency with our previous works, in the rest of this paper, we adopt the same notation already introduced in \cite{Sem_eff_est_TSP, For_MLSP}. In addition to the list of symbols detailed in \cite{Sem_eff_est_TSP}, we will make extensive use of some specific matrices whose definitions are collected below. In particular:
\be
\label{mat_P}
\mb{P} = \quadre{\mb{e}_2|\mb{e}_3|\cdots| \mb{e}_{N^2}}^\top,
\ee
where $\mb{e}_i$ is the $i$-th vector of the canonical basis of $\mathbb{R}^{N^2}$, the projection matrix
\be
\label{mat_proj_I}
\Pi^{\perp}_{\cvec{\mb{I}_N}}=\mb{I}_{N^2} - N^{-1}\mathrm{vec}(\mb{I}_N)\mathrm{vec}(\mb{I}_N)^\top,
\ee
and
\be
\label{L_mat}
\mb{L}_{\mb{V}_1} = \mb{P} \tonde{\kronVtinvmT} \Pi^{\perp}_{\cvec{\mb{I}_N}},
\ee
where $\mb{V}_1$ is the shape matrix previously introduced.

\textit{CES-related notation}: Without any claim of completeness, we collect below the basic properties and notation on CES distributed random vectors. We refer the reader to the excellent review paper \cite{Esa} for additional material. Let $\bs{\theta}_0 \triangleq (\bs{\mu}_0^\top,\bs{\mu}_0^\topH,\vc{\mb{V}_{1,0}}^\top)^\top$ be the \virg{true} parameter vector to be estimated and let $h_0$ be the actual (and unknown) density generator. Let $\mathbb{C}^N \ni \mb{z} \sim p_0(\mb{z}) \equiv p_Z(\mb{z};\bs{\theta}_0,h_0) \equiv CES_N(\bs{\mu}_0, \mb{V}_{1,0}, h_0)$ a CES-distributed random vector parameterized by a location vector $\bs{\mu}_0$, a shape matrix $\mb{V}_{1,0}\triangleq \bs{\Sigma}_0/[\bs{\Sigma}_0]_{1,1}$ where $\bs{\Sigma}_0$ represents the relevant scatter matrix and a density generator $h_0 \in \mathcal{G}$. Then, $\mb{z}$ satisfies the following stochastic representation:
\be
\mb{z} =_d \bs{\mu}_0 + \sqrt{\mathcal{Q}}\bs{\Sigma}_0^{1/2}\mb{u},
\ee
where $\mb{u} \sim \mathcal{U}(\mathbb{C}S^N)$ is a complex random vector uniformly distributed on the unit $N$-sphere and $=_d$ stands for \virg{has the
	same distribution as}. The \textit{2nd-order modular variate} $\mathcal{Q}$ is independent from $\mb{u}$ and such that (s.t.):
\be
\label{Q_CES}
\mathcal{Q}=_d (\mb{z}-\bs{\mu}_0)^H\bs{\Sigma}_0^{-1}(\mb{z}-\bs{\mu}_0)\triangleq Q,
\ee
Moreover, $\mathcal{Q}$ is distributed according to the following pdf:
\be
\label{pdf_Q}
p_{\mathcal{Q},0}(q) = \pi^{N}\Gamma(N)^{-1} q^{N-1} h_0 (q),
\ee
where $\Gamma(\cdot)$ is the Gamma function. For any $h \in \mathcal{G}$, the function $\psi$ is defined as
\be
\label{psi}
\psi(t) \triangleq d \ln h(t)/dt.
\ee
Finally, the expectation operator of any \textit{measurable} function $f$ with respect to $p_0(\mb{z})$ is indicated as $E_0\{f(\mb{z})\} \triangleq \int f(\mb{z}) p_Z(\mb{z};\bs{\theta}_0,h_0) d\mb{z}$.

\textit{Ranks}: The concept of \textit{ranks} of a set of relevant random variables are a useful tool in non-parametric statistics and numerous works can be found on this topic (see \cite{hajek1968}, \cite[Ch. 13]{vaart_1998} and references therein). Far be it from us to propose a comprehensive overview of the use of ranks in robust statistics, in the following we limit ourselves to introduce their definition since they will play a crucial role in the definition of the $R$-estimator of the shape matrix. Let $\{a_1,a_2,\ldots,a_L\}$ be a set of $L$ continuous i.i.d. random variables with unspecified distribution. Let us rearrange the variables $a_l$, $l=1,\ldots,L$ in an ascending order $a_{L(1)}<a_{L(2)}< \cdots < a_{L(L)}$ and, consequently build the vector of \textit{order statistics} as $\mb{v}_A \triangleq [a_{L(1)}, a_{L(2)},\ldots,a_{L(L)}]^\top$. Then, the \text{rank} $r_l \in \mathbb{N}/\{0\}$ of $a_l$ is the position index of $a_l$ in $\mb{v}_A$.

\section{A computationally efficient implementation of the $R$-estimator for shape matrices}
\label{sec_comp_eff_R}
Building upon the the seminal work of Hallin, Oja and Paindaveine \cite{Hallin_Annals_Stat_2}, in our recent papers \cite{Sem_eff_est_TSP, EUSIPCO_2020} a robust and semiparametric efficient $R$-estimator $\widehat{\mb{V}}_{1,R}$ for the shape matrix $\mb{V}_{1,0}$ of CES distributed data has been proposed and its properties investigated. This $R$-estimator has its roots in the Le Cam's theory of efficient \virg{one-step} estimator \cite{LeCam} and consequently it can be expressed as a linear combination of two terms: 
\be
\label{R_est_int}
\ovec{\widehat{\mb{V}}_{1,R}}  = \ovec{\widehat{\mb{V}}_1^\star} + L^{-1/2}\widehat{\bs{\Upsilon}_\mathbb{C}}^{-1}\widetilde{\bs{\Delta}}^\mathbb{C}_{\widehat{\mb{V}}_1^\star},
\ee
where $\pV$ is a $\sqrt{L}$-consistent preliminary estimator that provides $\widehat{\mb{V}}_{1,R}$ with the consistency property while the linear correction term $L^{-1/2}\widehat{\bs{\Upsilon}_\mathbb{C}}^{-1}\widetilde{\bs{\Delta}}^\mathbb{C}_{\widehat{\mb{V}}_1^\star}$ makes $\widehat{\mb{V}}_{1,R}$ semiparametric efficient. Let us have a closer look at the two quantities which constitute the linear correction term (all the details can be found in \cite{Sem_eff_est_TSP}). 

The $(N^2-1)$-dimensional vector $\widetilde{\bs{\Delta}}^\mathbb{C}_{\widehat{\mb{V}}_1^\star}$ is the \virg{distribution-free} version of the \textit{efficient central sequence} \cite{BKRW} and it can be explicitly expressed as:
\be\label{complex_app_eff_cs_1}
\widetilde{\bs{\Delta}}_{\widehat{\mb{V}}_1^\star}^{\mathbb{C}} \triangleq \frac{1}{\sqrt{L}}\mb{L}_{\widehat{\mb{V}}_1^\star}\sum_{l=1}^{L}K_h\tonde{\frac{r_l^\star}{L+1}}  \mathrm{vec}(\hat{\mb{u}}^\star_l(\hat{\mb{u}}^\star_l)^\topH),
\ee 
where $\{r_l^\star\}_{l=1}^L$ are the ranks of the random variables $\{\hat{Q}_l^\star\}_{l=1}^L$ defined as:
\be\label{CES_Q_star_1}
\hat{Q}^\star_l \triangleq (\mb{z}_l-\widehat{\bs{\mu}}^\star)^\mathsf{H}[\widehat{\mb{V}}^\star_1]^{-1}(\mb{z}_l-\widehat{\bs{\mu}}^\star),
\ee
$\widehat{\bs{\mu}}^\star$ and $\widehat{\mb{V}}^\star_1$ are two $\sqrt{L}$-consistent preliminary estimators \footnote{The choice of these preliminary estimators and of their impact on the asymptotic performance of $\widehat{\mb{V}}_{1,R}$ will be extensively discussed in the next Sections.} of the location vector $\bs{\mu}_0$ and of the shape matrix $\mb{V}_{1,0}$. The random vectors $\{\hat{\mb{u}}^\star_l\}_{l=1}^L$ are given by:
\be\label{CES_u_star_1}
\hat{\mb{u}}^\star_l \triangleq (\hat{Q}^\star_l)^{-1/2}[\widehat{\mb{V}}^\star_1]^{-1/2}(\mb{z}_l-\widehat{\bs{\mu}}^\star).
\ee
The function $K_h:(0,1)\rightarrow \mathbb{R}^+$ is the so-called \textit{score function} and it is a key element to guarantee the robustness of $\widehat{\mb{V}}_{1,R}$. We refer the reader to \cite{Sem_eff_est_TSP} for further details on the assumptions that a score function has to satisfy and on how to built it starting from the set of density generators $\mathcal{G}$.

The $(N^2-1) \times (N^2-1)$ matrix $\widehat{\bs{\Upsilon}_\mathbb{C}}$ represents the \virg{distribution-free} approximation of the \textit{semiparametric Fisher Information Matrix} (SFIM) and it is given by \cite[Eq. (52)]{Sem_eff_est_TSP}:
\be\label{Ups_mat}
\widehat{\bs{\Upsilon}} \triangleq \hat{\alpha}_\mathbb{C}\mb{L}_{\widehat{\mb{V}}_1^\star} \mb{L}_{\widehat{\mb{V}}_1^\star}^\topH,
\ee 
where $\hat{\alpha}_\mathbb{C}$ is a complex scalar that can be obtained as \cite[Eq. (53)]{Sem_eff_est_TSP}:
\be\label{com_alpha_hat_1}
\hat{\alpha}_\mathbb{C} = \frac{\norm{\widetilde{\bs{\Delta}}^\mathbb{C}_{\widehat{\mb{V}}_1^\star + L^{-1/2}\mb{H}^0_\mathbb{C}} - \widetilde{\bs{\Delta}}^\mathbb{C}_{\widehat{\mb{V}}_1^\star}}}{ \norm{ \mb{L}_{\widehat{\mb{V}}_1^\star} \mb{L}_{\widehat{\mb{V}}_1^\star}^\mathsf{H}\ovec{\mb{H}^0_\mathbb{C}}} },
\ee
and $\mb{H}^0_\mathbb{C}$ is a \virg{small perturbation}, Hermitian, matrix s. t. $[\mb{H}^0_\mathbb{C}]_{1,1}=0$. 

By substituting Eqs. \eqref{complex_app_eff_cs_1} and \eqref{Ups_mat} in Eq. \eqref{R_est_int}, the $R$-estimator $\widehat{\mb{V}}_{1,R}$ can be explicitly re-written as (see \cite[Eq. (54)]{Sem_eff_est_TSP}):
\be
\label{R_est}
\begin{split}
	\ovec{\widehat{\mb{V}}_{1,R}}  =& \ovec{\widehat{\mb{V}}_1^\star} + \\
	& \frac{1}{L\hat{\alpha}_\mathbb{C}}\quadre{\mb{L}_{\widehat{\mb{V}}_1^\star} \mb{L}_{\widehat{\mb{V}}_1^\star}^\mathsf{H}}^{-1}
	\mb{L}_{\widehat{\mb{V}}_1^\star}\sum\nolimits_{l=1}^{L}K_h\tonde{\frac{r_l^\star}{L+1}} \mathrm{vec}(\hat{\mb{u}}^\star_l(\hat{\mb{u}}^\star_l)^\mathsf{H}).
\end{split}
\ee

For an in-depth discussion about the semiparametric efficiency and the robustness property characterizing $\widehat{\mb{V}}_{1,R}$, we refer the readers to our previous works \cite{Sem_eff_est_TSP, EUSIPCO_2020} and to the related statistical literature \cite{Hallin_P_Annals,Hallin_Annals_Stat_2,Hallin_P_2006}. Here we focus our attention on an important aspect that has not been fully addressed yet: the computational cost underlying the calculation of Eq. \eqref{R_est}. As already noted in \cite[Sec. V.C]{Sem_eff_est_TSP}, there is a main (computational) drawback in Eqs. \eqref{R_est} and \eqref{com_alpha_hat_1} that really stands out: to evaluate the $N \times N$ matrix $\widehat{\mb{V}}_{1,R}$ (or equivalently its $(N^2-1)$-dimensional vectorized counterpart $\ovec{\widehat{\mb{V}}_{1,R}}$), we have to calculate the $(N^2-1)\times (N^2-1)$ matrix $\mb{L}_{\widehat{\mb{V}}_1^\star}$. This may become a cumbersome bottleneck in many practical applications.

Fortunately, as proved in Appendix A of this paper, it is possible to recast Eqs. \eqref{R_est} and \eqref{com_alpha_hat_1} in order to avoid the calculation of $\mb{L}_{\widehat{\mb{V}}_1^\star}$. In particular, a computationally efficient \virg{matrix version} of the $R$-estimator $\widehat{\mb{V}}_{1,R}$ can be expressed as:
\be
\label{mat_one_step_R_new_1}
\boxed{\widehat{\mb{V}}_{1,R} = \widehat{\mb{V}}_1^\star +\frac{1}{\hat{\alpha}_\mathbb{C}} \tonde{\mb{W} - \quadre{\mb{W}}_{1,1}\widehat{\mb{V}}_1^\star}},
\ee
\be\label{com_alpha_hat_comp_1}
\boxed{\hat{\alpha}_\mathbb{C} = \frac{\norm{\underline{\mb{z}}_{\widehat{\mb{V}}_1^\star + L^{-1/2}\mb{H}^0_\mathbb{C}} - \underline{\mb{z}}_{\widehat{\mb{V}}_1^\star}}}{\left\|  \underline{\mathrm{vec}}\tonde{(\pV)^{-1} \mb{H}^0_\mathbb{C} (\pV)^{-1} -N^{-1} \mathrm{tr}\tonde{(\pV)^{-1} \mb{H}^0_\mathbb{C}} (\pV)^{-1}}\right\| }},
\ee
where:
\be\label{mat_W_1}
\mb{W} \triangleq L^{-1/2}(\pV)^{1/2} \mb{R} (\pV)^{1/2}.
\ee
\be\label{Z_star_1}
\underline{\mb{z}}_{\widehat{\mb{V}}_1^\star} \triangleq \underline{\mathrm{vec}}\tonde{(\pV)^{-1/2} \mb{R} (\pV)^{-1/2} - \zeta (\pV)^{-1}},
\ee
\be\label{def_R_1}
\mb{R} \triangleq \frac{1}{\sqrt{L}}\sum_{l=1}^{L}K_h\tonde{\frac{r_l^\star}{L+1}}\hat{\mb{u}}^\star_l(\hat{\mb{u}}^\star_l)^\topH,
\ee
\be\label{def_zeta_1} 
\zeta \triangleq  \frac{1}{N\sqrt{L}}\sum_{l=1}^{L}K_h\tonde{\frac{r_l^\star}{L+1}}.
\ee

It is worth stressing that Eqs. \eqref{mat_one_step_R_new_1} and \eqref{com_alpha_hat_comp_1} involve matrix and vector quantities of (linear) dimension equal at most to $N^2$ and this fact leads to a great reduction in terms of computational load with respect to Eqs. \eqref{R_est} and \eqref{com_alpha_hat_1} that, on the contrary, rely on the calculation of matrices whose linear dimension is $N^4$. A quantitative analysis of the reduction of the computational load will be provided in Sec. \ref{sec:num}. We conclude this Section by noticing that an expression similar to Eq. \eqref{mat_one_step_R_new_1} characterizing the $R$-estimator of the shape matrix of a set of Real Elliptically Symmetric (RES) distributed data has been firstly provided in \cite[Eq. (3.9)]{Hallin_Annals_Stat_2}. 

\textit{Remark}: The interested reader can find our Matlab implementation of the computationally efficient $R$-estimator $\widehat{\mb{V}}_{1,R}$ for the shape matrix of both Real and Complex Elliptically Symmetric distributed data at 
\url{https://github.com/StefanoFor}. 
 
\section{Semiparametric joint estimation of location and shape: the role of the nuisance density generator}
\label{joint_est}

After having introduced a computationally efficient version of the $R$-estimator of the shape matrix $\mb{V}_{1,0}$, in this Section we will focus on a different still interrelated topic: which is the impact of not knowing the location vector $\bs{\mu}_0$ when estimating $\mb{V}_{1,0}$? Along with its theoretical implication, the answer to this question has a practical importance as well. As shown in the previous Section in fact (see Eqs. \eqref{CES_Q_star_1} and \eqref{CES_u_star_1}), the $R$-estimator $\widehat{\mb{V}}_{1,R}$ relies on $\widehat{\bs{\mu}}^\star$, i.e. a $\sqrt{L}$-consistent preliminary estimator of $\bs{\mu}_0$. However, Section \ref{sec_comp_eff_R} does not provide any suggestion on which specific estimator $\widehat{\bs{\mu}}^\star$ should we choose among all the possible $\sqrt{L}$-consistent ones. In this Section then we are going to provide with the necessary theoretical framework that will allow us to make the good choice for $\widehat{\bs{\mu}}^\star$.     

Let us start by formalizing the problem. Let $\{\mb{z}_l\}_{l=1}^L$ be a set of CES distributed vectors such that $\mathbb{C}^N \ni \mb{z}_l \sim p_0 \equiv CES_N(\bs{\mu}_0, \mb{V}_{1,0}, h_0)$, $\forall l$ where $\bs{\mu}_0$ and $\mb{V}_{1,0}$ have to be considered as two finite-dimensional parameters of interest while $h_0$ is a functional nuisance term. 

The two fundamental questions underlying the above mentioned joint estimation problem are:
\begin{enumerate}
	\item What is the impact of not knowing $h_0$ on the joint estimation of $(\bs{\mu}_0, \mb{V}_{1,0})$?
	\item What is the (asymptotic) impact that the lack of knowledge of $\bs{\mu}_0$ has on the estimation of $\mb{V}_{1,0}$ and vice versa?
\end{enumerate}

To answer these two questions, we need to introduce the \textit{semiparametric efficient score vector} $\bar{\mb{s}}_{\bs{\theta}_0}$ and the \textit{semiparamatric Fisher Information Martix} (SFIM) $\bar{\mb{I}}(\bs{\theta}_0|h_0)$. As discussed in the relevant statistical literature for a generic semiparametric model \cite{BKRW} and recently investigated for the specific CES model \cite{For_SCRB, For_SCRB_complex}, the semiparametric efficient score vector for the joint estimation of location and shape matrix of a set of CES distributed data is given by:
\be
\label{eff_com_scor_vect}
\bar{\mb{s}}_{\bs{\theta}_0} = [\bar{\mb{s}}^\top_{\bs{\mu}_0},\bar{\mb{s}}^\top_{\bs{\mu}^*_0},\bar{\mb{s}}^\top_{\underline{\mathrm{vec}}(\mb{V}_{1,0})}]^\top = \mb{s}_{\bs{\theta}_0} - \Pi(\mb{s}_{\bs{\theta}_0}|\mathcal{T}_{h_0}),
\ee 
where $\mb{s}_{\bs{\theta}_0}$ is the \virg{classical} score vector defined, by means of the Wirtinger derivatives, as \cite{Complex_M}:
\be
[\mb{s}_{\bs{\theta}_0}]_i \triangleq \left. \partial\ln p_Z(\mb{z};\bs{\theta},h_0)/\partial\theta_i^*\right|_{\bs{\theta}=\bs{\theta}_0},\; i=1,\ldots,q
\ee 
and $\bs{\theta}$ is given in Eq. \eqref{theta_com} and $q=N(N+2)-1$. The term $\Pi(\mb{s}_{\bs{\theta}_0}|\mathcal{T}_{h_0})$ indicates the orthogonal projection of $\mb{s}_{\bs{\theta}_0}$ on the nuisance tangent space $\mathcal{T}_{h_0}$ of the CES model $\mathcal{P}_{\bs{\theta},h}$ in Eq. \eqref{CES_semi_par_model} evaluated at the true density generator $h_0$. Specifically, $\Pi(\mb{s}_{\bs{\theta}_0}|\mathcal{T}_{h_0})$ tells us the loss of information on the estimation of $\bs{\theta}_0$ due to the lack of knowledge of $h_0$. In our previous work \cite[Sec. III.A]{For_SCRB_complex}, we proved the following facts:
\begin{enumerate}
	\item The projection of $\bar{\mb{s}}_{\bs{\mu}_0}$ and $\bar{\mb{s}}_{\bs{\mu}^*_0}$ onto $\mathcal{T}_{h_0}$ is equal to zero:  
	\be
	\label{proj_mu}
	\Pi(\mb{s}_{\bs{\mu}_0}|\mathcal{T}_{h_0}) = \Pi(\mb{s}_{\bs{\mu}^*_0}|\mathcal{T}_{h_0}) = \mb{0}_N,
	\ee
	This implies that the lack of knowledge of $h_0$ does not have any impact on the (asymptotic) estimation of the location parameter $\bs{\mu}_0$.
	\item The projection of $\bar{\mb{s}}_{\underline{\mathrm{vec}}(\mb{V}_{1,0})}$ onto $\mathcal{T}_{h_0}$ is generally different from zero and it is given by:
	\be
	\label{proj_vect}
	\Pi(\mb{s}_{\underline{\mathrm{vec}}(\mb{V}_{1,0})}|\mathcal{T}_{h_0}) = -(1+N^{-1}\mathcal{Q}\psi_0(\mathcal{Q}))\ovec{\mb{V}_{1,0}^{-1}},
	\ee
	where $\psi_0$ is defined in Eq. \eqref{psi} while $\mathcal{Q}$ is given in Eq. \eqref{Q_CES}. Consequently, not knowing $h_0$ does have an impact on the estimation of the shape matrix $\mb{V}_{1,0}$.
\end{enumerate}

Points 1) and 2) answer the first question.

To address the second question about the (asymptotic) cross-information between $\bs{\mu}_0$ and $\mb{V}_{1,0}$, we need to check the structure of the SFIM $\bar{\mb{I}}(\bs{\theta}_0|h_0)$. The SFIM for the joint estimation of $\bs{\mu}_0$ and $\mb{V}_{1,0}$ in the CES semiparametric model $\mathcal{P}_{\bs{\theta},h}$ in Eq. \eqref{CES_semi_par_model} has been evaluated in \cite[Sec. III.C]{For_SCRB_complex} as:
\be
\label{SFIM_L}
	\bar{\mb{I}}(\bs{\theta}_0|h_0) \triangleq E_0\{\bar{\mb{s}}_{\bs{\theta}_0}\bar{\mb{s}}_{\bs{\theta}_0}^\topH\} = \left(
	\begin{array}{cc}
		\bar{\mb{I}}(\bs{\mu}_0|h_0) & \mb{0}_{2N \times (N^2-1)}\\
		\mb{0}_{(N^2-1) \times 2N} & \bar{\mb{I}}(\mb{V}_{1,0}|h_0)
	\end{array}
	\right),
\ee
\be
\label{SFIM_com_mu}
\bar{\mb{I}}(\bs{\mu}_0|h_0) = \frac{E\{\mathcal{Q}\psi_0(\mathcal{Q})^2\}}{N}	\left( \begin{array}{cc}
	\mb{V}_{1,0}^{-1} & \mb{0}_{N \times N}\\
	\mb{0}_{N \times N} & \mb{V}_{1,0}^{-*}
\end{array}\right),
\ee
\be
\label{com_cov_mat_eff_scat}
\begin{split}
	\bar{\mb{I}}(&{\ovec{\mb{V}_{1,0}}}|h_0) = \frac{E\{\mathcal{Q}^2\psi_0(\mathcal{Q})^2\}}{N(N+1)}\mb{L}_{\mb{V}_{1,0}} \mb{L}_{\mb{V}_{1,0}}^H = \\ &=\frac{E\{\mathcal{Q}^2\psi_0(\mathcal{Q})^2\}}{N(N+1)}\mb{P}\quadre{\mb{V}_{1,0}^{-\top}\otimes\mb{V}_{1,0}^{-1}-N^{-1}\vc{\mb{V}_{1,0}^{-1}}\vc{\mb{V}_{1,0}^{-1}}^\topH}\mb{P}^\top,
\end{split}
\ee   
where, as before, the function $\psi_0$ is given in Eq. \eqref{psi} while $\mathcal{Q}$ is given in Eq. \eqref{Q_CES}. Eq. \eqref{SFIM_L} clearly shows that the efficient SFIM $\bar{\mb{I}}(\bs{\theta}_0|h_0)$ is a block-diagonal matrix, i.e. the cross-information terms between the location $\bs{\mu}_0$ and the shape matrix $\mb{V}_{1,0}$ are equal to zero. Consequently, the relevant estimation problems are (asymptotically) decorrelated and can be considered as two separate estimation problems. This fact greatly simplify the implementation of a practical joint estimation algorithm. In fact, in estimating the shape matrix $\mb{V}_{1,0}$, the true (and generally unknown) location vector $\bs{\mu}_0$ can be substituted by any of its $\sqrt{L}$-consistent estimators without any impact on the (asymptotic) performance of the estimator of $\mb{V}_{1,0}$. Of course, the vice versa holds true as well, i.e.\ any $\sqrt{L}$-consistent estimator of $\mb{V}_{1,0}$ can be used in place of the true shape matrix without any (asymptotic) impact on the estimation of $\bs{\mu}_0$. This important theoretical result will be exploited in the next Section, to implement a \textit{robust}, \textit{semiparametric efficient} joint estimator for the location and shape matrix in CES distributed data.

\section{A robust semiparametric efficient joint estimator of location and shape}
\label{rob_sem_eff_est}
Robust estimation of location and shape in elliptical distributions is a well-known topic in statistics and signal processing since the seminal paper of Maronna \cite{maronna1976}. In particular, in \cite{maronna1976}, a general class of joint $M$-estimators of $\bs{\mu}_0$ and $\mb{V}_{1,0}$ (in the presence of an unknown density generator $h_0$) has been introduced as the \virg{fixed-point} solution of the following system of equations:
\be
\label{M_mu}
\sum\nolimits_{l=1}^{L}u_1(\hat{Q}_l^{1/2})(\mb{z}_l-\hat{\bs{\mu}})=\mb{0},
\ee
\be
\label{M_shape}
L^{-1} \sum\nolimits_{l=1}^{L}u_2(\hat{Q}_l)(\mb{z}_l-\hat{\bs{\mu}})(\mb{z}_l-\hat{\bs{\mu}})^\topH=\widehat{\mb{V}}_{1},
\ee
where, according to Eq. \eqref{Q_CES}, $\hat{Q}_l = (\mb{z}_l-\hat{\bs{\mu}})^\topH \widehat{\mb{V}}_{1}^{-1}(\mb{z}_l-\hat{\bs{\mu}})$, and $\{\mb{z}_l\}_{l=1}^L$ is the set of available CES distributed observations such that $\mathbb{C}^N \ni \mb{z}_l \sim p_0 \equiv CES_N(\bs{\mu}_0, \mb{V}_{1,0}, h_0)$, $\forall l$. The functions $u_1$ and $u_2$ have to satisfy a given set of assumptions that guarantees the existence and the uniqueness of the solution of Eqs. \eqref{M_mu} and \eqref{M_shape} (see \cite{maronna1976} for the real case and \cite{Esa} for the extension to the complex one). 

\subsection{Tyler's joint $M$-estimator of $\bs{\mu}_0$ and $\mb{V}_{1,0}$}
\label{sec_Tyler}
Among different possible choices for $u_1$ and $u_2$, Tyler in \cite{Tyler1} (see also \cite{Joana}, \cite{Meriaux_joint} and \cite{violeta}) showed that the functions $u_1(Q)=Q^{-1/2}$ and $u_2(Q)=NQ^{-1}$ lead to the \virg{minimax robust} $M$-estimator of the location and shape. Specifically, by defining 
\be
\hat{Q}_l^{(k)} = (\mb{z}_l-\hat{\bs{\mu}}^{(k)})^\topH [\widehat{\mb{V}}_{1}^{(k)}]^{-1}(\mb{z}_l-\hat{\bs{\mu}}^{(k)}),
\ee 
where $k$ indicates the iteration number, we have that the Tyler's joint $M$-estimator of location and shape, i.e.\ $(\hat{\bs{\mu}}_{Ty},\widehat{\mb{V}}_{1,Ty})$, can be obtained as the convergence points ($k\rightarrow \infty$) of the following iterations:
\be
\label{Ty_mu}
\hat{\bs{\mu}}_{Ty}^{(k+1)} =\quadre{\sum_{l=1}^{L}[\hat{Q}_l^{(k)}]^{-1/2}}^{-1}\sum_{l=1}^{L}\tonde{\hat{Q}_l^{(k)}}^{-1/2}\mb{z}_l,
\ee   
\be
\label{Ty_schape}
\left\lbrace  \begin{array}{l}
	\widehat{\mb{V}}_{Ty}^{(k+1)} =\frac{N}{L}\sum\limits_{l=1}^{L}\frac{(\mb{z}_l-\hat{\bs{\mu}}_{Ty}^{(k)})(\mb{z}_l-\hat{\bs{\mu}}_{Ty}^{(k)})^\topH}{\hat{Q}_l^{(k)}}.\\
	\widehat{\mb{V}}_{1,Ty}^{(k+1)} \triangleq \nicefrac{\widehat{\mb{V}}_{Ty}^{(k+1)}}{[\widehat{\mb{V}}_{Ty}^{(k+1)}]_{1,1}}.
\end{array}\right. 
\ee   
Note that, even if a formal proof of the joint convergence of Eqs. \eqref{Ty_mu} and \eqref{Ty_schape} is still an open problem, this iterative algorithm has been shown to provide reliable estimates in most of the scenarios of possible interest in practical applications. We refer to \cite{Joana} where joint $M$-estimators of the form \eqref{M_mu}-\eqref{M_shape} have been exploited in hyperspectral anomaly detection problems and to \cite{violeta} where joint $M$-estimators have been derived as part of a general Expectation-Maximization (EM) algorithm for clustering applications.   

The estimators $\hat{\bs{\mu}}_{Ty}$ and $\widehat{\mb{V}}_{1,Ty}$ have the remarkable property of being $\sqrt{L}$-consistent under any (unknown) density generator $h \in \mathcal{G}$ (see \cite{Tyler1} for the real-valued case and \cite{Meriaux_joint} for the complex-valued case). Consistency, however, is only one of the properties that good robust estimators should have. Another important property is the (semiparametric) efficiency. 
\subsection{The Semiparametric Cram\'er-Rao Bound (SCRB)}
A robust estimator is said to be \textit{semiparametric efficient} if its Mean Square Error (MSE) achieves the Semiparametric Cram\'er-Rao Bound (SCRB) \cite{BKRW} as the number of available observations $L$ goes to infinity. The SCRB for the joint estimation of location and shape in CES distributed data has been derived in \cite{For_SCRB_complex} as the inverse of the SFIM in Eq. \eqref{SFIM_L}. We refer the reader to \cite{For_SCRB_complex} for all the details about its calculation. Here, for the sake of conciseness, we report only the final expression. As discussed before, since the efficient score vectors for the location, i.e.\  $\bar{\mb{s}}_{\bs{\mu}_0}$ and $\bar{\mb{s}}_{\bs{\mu}^*_0}$, are orthogonal to the nuisance tangent space $\mathcal{T}_{h_0}$, the SCRB on the estimation of $\bs{\mu}_0$ is equal to the \virg{classical} CRB and it is given by
\be
\label{SCRB_mu}
\begin{split}
	\mathrm{SCRB}(\bs{\mu}_0|h_0)  \triangleq \bar{\mb{I}}(\bs{\mu}_0|h_0)^{-1} = \frac{N}{E\{\mathcal{Q}\psi_0(\mathcal{Q})^2\}}	\left( \begin{array}{cc}
		\mb{V}_{1,0} & \mb{0}_{N \times N}\\
		\mb{0}_{N \times N} & \mb{V}_{1,0}^*
	\end{array}\right).
\end{split}
\ee
On the other hand, since as previously shown in Eq. \eqref{proj_vect}, $\Pi(\mb{s}_{\ovec{\mb{V}_{1,0}}}|\mathcal{T}_{h_0}) \neq \mb{0}$, the SCRB on the estimation of the shape matrix $\mb{V}_{1,0}$ is tighter than the \virg{classical} CRB (that is obtained for a perfectly known $h_0$) and is given by:
\be
\label{SCRB_shape}
\mathrm{SCRB}({\ovec{\mb{V}_{1,0}}}|h_0)\triangleq \bar{\mb{I}}({\ovec{\mb{V}_{1,0}}}|h_0)^{-1} = \frac{N(N+1)}{E\{\mathcal{Q}^2\psi_0(\mathcal{Q})^2\}} \quadre{\mb{L}_{\mb{V}_{1,0}} \mb{L}_{\mb{V}_{1,0}}^H}^{-1},
\ee  
where the matrix $\mb{L}_{\mb{V}_{1,0}}$ is defined in Eq. \eqref{L_mat}.
It is worth mentioning that the expression of the SCRB given in Eq. \eqref{SCRB_shape} is valid only if the shape matrix is defined through the constraint in Eq. \eqref{const_11}, i.e. when the first-top left element of $\mb{V}_{1,0}$ is forced to be equal to 1. The interested reader may find the general form of the SCRB for the shape matrix estimation under any constraints (e.g. constraints on its trace or determinant) in \cite{For_SCRB,For_SCRB_complex}.

In \cite{For_SCRB,For_SCRB_complex}, it has been shown that robust $M$-estimators of the shape matrix are not semiparametric efficient. This efficiency issue can be overcome by exploiting the $R$-estimator of the shape given in Eqs. \eqref{mat_one_step_R_new_1} in Sec. \ref{sec_comp_eff_R}.

\subsection{An $R$-estimator of $\mb{V}_{1,0}$ in non-centered CES data}
In this subsection, we finally put all our previous results together to provide a robust and semiparametric efficient joint estimation of the location vector $\bs{\mu}_0$ and of the shape matrix $\mb{V}_{1,0}$ of a set $\{\mb{z}_l\}_{l=1}^L$ of CES-distributed observations. As previously discussed, in order to gain the semiparametric efficiency, we will exploit the $R$-estimator in Eq. \eqref{mat_one_step_R_new_1}. As amply discussed in Section \ref{sec_comp_eff_R}, to implement this estimator we need:
\begin{enumerate}
	\item A preliminary $\sqrt{L}$-consistent estimator $\widehat{\mb{V}}^\star_1$ for the shape matrix and another $\sqrt{L}$-consistent estimator of the location vector $\hat{\bs{\mu}}^\star$,
	\item A \text{score function} $K_h: (0,1) \rightarrow \mathbb{R}^+$. 
\end{enumerate}   

Due to their properties of minimax robustness and $\sqrt{L}$-consistency under any density generator $h \in \mathcal{G}$, the Tyler's estimators previously introduced in Eqs. \eqref{Ty_mu} and \eqref{Ty_schape} are perfect candidates for this role, i.e. $\hat{\bs{\mu}}^\star \equiv \hat{\bs{\mu}}_{Ty}$ and $\widehat{\mb{V}}^\star_1 \equiv \widehat{\mb{V}}_{1,Ty}$. Specifically, the general expression of the (computationally efficient) $R$-estimator given in Section \ref{sec_comp_eff_R} can be recast as:
\be
\label{mat_one_step_R_new_Ty}
\widehat{\mb{V}}_{1,R} = \widehat{\mb{V}}_{1,Ty} +\frac{1}{\hat{\alpha}_\mathbb{C}} \tonde{\mb{W} - \quadre{\mb{W}}_{1,1}\widehat{\mb{V}}_{1,Ty}}.
\ee
Note that, all the other related quantities reported in Eqs. \eqref{mat_W_1} - \eqref{def_zeta_1} have to be evaluated by substituting to the generic preliminary estimators $\hat{\bs{\mu}}^\star$ and $\widehat{\mb{V}}^\star_1$ with the Tyler's estimators for location and scale, $\hat{\bs{\mu}}_{Ty}$ and $\widehat{\mb{V}}_{1,Ty}$ respectively.

Regarding the second point, i.e. the choice of a score function $K_h$, we will exploit two different options \cite{Sem_eff_est_TSP}:
\begin{itemize}
	\item The complex \textit{van der Waerden} score function:
	\be\label{K_van}
	K_{vdW}(u) \triangleq \Phi_G^{-1}(u),\quad u \in (0,1),
	\ee
	where $\Phi_G^{-1}$ indicates the inverse function of the cdf of a Gamma-distributed random variable with parameters $(N,1)$.
	\item The complex $t_{\nu}$-score given by:
	\be\label{K_tnu}
	K_{t_\nu}(u) = \frac{N(2N+\nu)F^{-1}_{2N,\nu}(u)}{\nu + 2NF^{-1}_{2N,\nu}(u)},\quad u \in (0,1),
	\ee
	where $F_{2N,\nu}(u)$ stands for the Fisher cdf with $2N$ and $\nu \in (0,\infty)$ degrees of freedom.
\end{itemize}

The \textit{van der Waerden} score $K_{vdW}$ has been proved to have excellent performance in terms of efficiency in the estimation of the shape matrix in centered CES data \cite{PAINDAVEINE_CS, Sem_eff_est_TSP} while the $t_{\nu}$-score $K_{t_\nu}$ is able to provide with a better robustness to the presence of possible outliers thanks to the presence of the additional \virg{tuning} parameter $\nu$.     

To conclude this sub section, the pseudocode for the implementation of the $R$-estimator in Eq. \eqref{mat_one_step_R_new_Ty} is provided in the following while the related Matlab code can be downloaded at \url{https://github.com/StefanoFor}.

\begin{algorithm}
	\caption{Computationally efficient $R$-estimator for $\mb{V}_{1,0}$}
	\begin{algorithmic}[1]
		\renewcommand{\algorithmicrequire}{\textbf{Input:}}
		\renewcommand{\algorithmicensure}{\textbf{Output:}}
		\REQUIRE $\mb{z}_1,\ldots,\mb{z}_L$.
		\ENSURE  $\widehat{\mb{V}}_{1,R}$.
		\STATE Evaluate the preliminary Tyler's joint estimators:\\
		$\hat{\bs{\mu}}_{Ty} \leftarrow \lim_{k\rightarrow \infty}\hat{\bs{\mu}}_{Ty}^{(k)}$ in \eqref{Ty_mu},\\
		$\widehat{\mb{V}}_{1,Ty} \leftarrow \lim_{k\rightarrow \infty} \widehat{\mb{V}}_{1,Ty}^{(k+1)}$ in \eqref{Ty_schape},
		\STATE Data centring: $\{\mb{z}_l\}_{l=1}^L \leftarrow \{\mb{z}_l-\hat{\bs{\mu}}_{Ty}\}_{l=1}^L$,
		\FOR {$l = l$ to $L$} 
		\STATE $\hat{Q}^\star_l \leftarrow \mb{z}_l^H\widehat{\mb{V}}_{1,Ty}^{-1}\mb{z}_l$,
		\STATE $\hat{\mb{u}}^\star_l \leftarrow (\hat{Q}^\star_l)^{-1/2}\widehat{\mb{V}}_{1,Ty}^{-1/2}\mb{z}_l$,
		\ENDFOR
		\STATE Evaluate the ranks $\{r_1^\star,\ldots,r_L^\star\}$ of $\{\hat{Q}^\star_1,\ldots,\hat{Q}^\star_L\}$,
		\STATE Select a score function $K_h$ ($K_{vdW}$ in \eqref{K_van} and $K_{t_\nu}$ in \eqref{K_tnu} are two options), 
		\STATE $\mb{R} \leftarrow  \frac{1}{\sqrt{L}}\sum_{l=1}^{L}K_h\tonde{\frac{r_l^\star}{L+1}}\hat{\mb{u}}^\star_l(\hat{\mb{u}}^\star_l)^\topH$,
		\STATE $\zeta \leftarrow  \frac{1}{N\sqrt{L}}\sum_{l=1}^{L}K_h\tonde{\frac{r_l^\star}{L+1}}$,
		\STATE $\mb{W} \leftarrow L^{-1/2}(\widehat{\mb{V}}_{1,Ty})^{1/2} \mb{R} (\widehat{\mb{V}}_{1,Ty})^{1/2}$,
		\STATE $\underline{\mb{z}}_{\widehat{\mb{V}}_{1,Ty}} \leftarrow \underline{\mathrm{vec}}\tonde{(\widehat{\mb{V}}_{1,Ty})^{-1/2} \mb{R} (\widehat{\mb{V}}_{1,Ty})^{-1/2} - \zeta (\widehat{\mb{V}}_{1,Ty})^{-1}},$
		\STATE Obtain $\hat{\alpha}_\mathbb{C}$ through the following two steps:\\
		1) Generate a random Hermitian matrix $\mb{H}^0_\mathbb{C}$ s.t. $[\mb{H}^0_\mathbb{C}]_{1,1}=0$, \\
		2) $\hat{\alpha}_\mathbb{C} \leftarrow \frac{\norm{\underline{\mb{z}}_{\widehat{\mb{V}}_{1,Ty} + L^{-1/2}\mb{H}^0_\mathbb{C}} - \underline{\mb{z}}_{\widehat{\mb{V}}_{1,Ty}}}}{\left\|  \underline{\mathrm{vec}}\tonde{(\widehat{\mb{V}}_{1,Ty})^{-1} \mb{H}^0_\mathbb{C} (\widehat{\mb{V}}_{1,Ty})^{-1} -N^{-1} \mathrm{tr}\tonde{(\widehat{\mb{V}}_{1,Ty})^{-1} \mb{H}^0_\mathbb{C}} (\widehat{\mb{V}}_{1,Ty})^{-1}}\right\| }$,\\
		\textit{Remark:} The entries of $\mb{H}^0_\mathbb{C}$ should be \virg{small enough} to guarantee that $\widehat{\mb{V}}_{1,Ty} + L^{-1/2}\mb{H}^0_\mathbb{C}$ is a positive definite matrix.
		\STATE The last step: $\widehat{\mb{V}}_{1,R} \leftarrow \widehat{\mb{V}}_{1,Ty} +\frac{1}{\hat{\alpha}_\mathbb{C}} \tonde{\mb{W} - \quadre{\mb{W}}_{1,1}\widehat{\mb{V}}_{1,Ty}}$,
		\RETURN $\widehat{\mb{V}}_{1,R}$ 
	\end{algorithmic} 
\end{algorithm}  

\section{Numerical results}
\label{sec:num}

This Section will be basically divided in two parts. In the first one, we discuss the computational advantages that the \virg{matrix version} of the $R$-estimator provided in Eq. \eqref{mat_one_step_R_new_1} has with respect to the \virg{vectorized version} derived in \cite{Sem_eff_est_TSP} and recalled in Eq. \eqref{R_est}. In the second part we finally assess, through numerical simulations, the semiparametric efficiency of the joint estimator $(\hat{\bs{\mu}}_{Ty},\widehat{\mb{V}}_{1,R})$, given in Eqs. \eqref{Ty_mu} and \eqref{mat_one_step_R_new_Ty}, respectively.

\textit{Data generation}: In both the two parts, we generate the set of non-centered CES distributed data $\{\mb{z}_l\}_{l=1}^L$ according to a Generalized Gaussian (GG) distribution \cite{Pascal_GG}, such that $\mathbb{C}^N \ni \mb{z}_l \sim p_0$, $\forall l$ where:
\be
p_0(\mb{z}_l) = |\bs{\Sigma}_0|^{-1}   h_0\left((\mb{z}_l-\bs{\mu}_0)^\topH\bs{\Sigma}_0^{-1}(\mb{z}_l-\bs{\mu}_0) \right), 
\ee
while the relevant density generator is given by:
\be
\label{GG_h}
h_0(t) \triangleq \frac{s\Gamma(N)b^{-N/s}}{\pi^{N}\Gamma({N/s)}} \exp \left( -\frac{t^s}{b}\right),\; t\in \mathbb{R}^+ .
\ee
We chose the GG distribution to assess the performance of the proposed joint estimator because of its flexibility in characterizing the data \virg{heavy-tailness} with respect to the Gaussian one. In fact, according to the value of the shape parameter $s>0$, the GG density generator in \eqref{GG_h} is able to define a distribution with both heavier tails ($0<s<1$) and lighter tails ($s>1$) compared to the Gaussian one ($s=1$). 

The parameters adopted in our simulations are:
\begin{itemize}
	\item $\bs{\Sigma}_0$ is a Toeplitz Hermitian matrix whose first column is given by $[1,\rho, \ldots,\rho^{N-1}]^T$; $\rho = 0.8e^{j2\pi/5}$ and $N=8$.
	\item Shape matrix: $\mb{V}_{1,0}\triangleq \bs{\Sigma}_0/[\bs{\Sigma}_0]_{1,1}$.
	\item Location vector: $[\bs{\mu}_0]_n \triangleq 0.5e^{j1\pi/7(n-1)}$, $n=1,\ldots,N$.
	\item Scale parameter: $b=[\sigma_X^2 N  \Gamma( N/s )/\Gamma( (N+1)/s ) ]^s$ in Eq. \eqref{GG_h} and $\sigma_X^2 = E\{\mathcal{Q}\}/N = 4$.
	\item Numbers of observations: $L=5N$. This clearly defines a \virg{finite-sample} regime.
\end{itemize}

\subsection{Computational efficiency of the proposed \virg{matrix version} of the $R$-estimator for shape}

\begin{figure}[h]
	\centering
	\includegraphics[height=5cm]{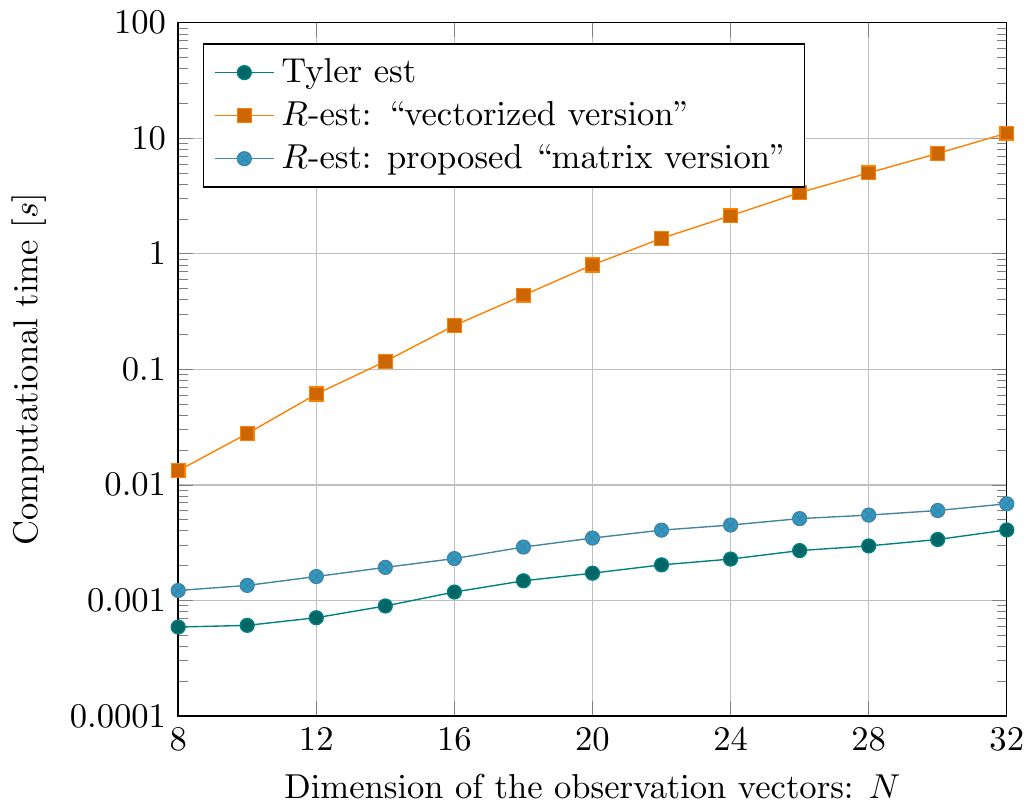}
	\caption{Comparison among the time riquired to calculate the Tyler's estimator, the \virg{vectorized} and the proposed \virg{matrix} versions of the $R$-estimator.}
	\label{fig:Fig1}
\end{figure}

As amply discussed in Section \ref{sec_comp_eff_R}, the crucial difference between the \virg{vectorized} implementation of the $R$-estimator given in Eq. \eqref{R_est} and its \virg{matrix version} provided in Eq. \eqref{mat_one_step_R_new_1} is in the fact that, while the first one relies on the calculation of $N^2 \times N^2$ matrix quantities, the latter only involves  $N \times N$ matrices. Clearly, this will lead to a huge gain in term of computational efficiency, in particular when the data dimension $N$ increases. In order to highlight this fact, in Fig. \ref{fig:Fig1} we report the time (in seconds) required for the calculation of three shape matrix estimators as function of the data dimension $N$:
\begin{itemize}
	\item The Tyler's estimator $\widehat{\mb{V}}_{1,Ty}$ in Eq. \eqref{Ty_schape},
	\item The \virg{vectorized version} of the $R$-estimator in Eq. \eqref{R_est} exploiting $\widehat{\mb{V}}_{1,Ty}$ as preliminary estimator for the shape matrix,
	\item The \virg{matrix version} of the $R$-estimator in Eq. \eqref{mat_one_step_R_new_1} exploiting $\widehat{\mb{V}}_{1,Ty}$ as preliminary estimator for the shape matrix.
\end{itemize} 

The curves in Fig. \ref{fig:Fig1} are crystal clear: the proposed \virg{matrix version} of the $R$-estimator is more than two orders of magnitude faster that the \virg{vectorized} one derived in \cite{Sem_eff_est_TSP}. The gap between the two clearly increases as the data dimension $N$ increases. Moreover, it can be noted that the computational time of the \virg{matrix version} of the $R$-estimator in Eq. \eqref{mat_one_step_R_new_1} is similar to the one of the Tyler's estimator. These considerations provide us with an hard evidence in favor of the computational effectiveness of the \virg{matrix version} of the $R$-estimator derived in Section \ref{sec_comp_eff_R} and suggest us to adopt it as standard form of the $R$-estimator.

\subsection{Statistical efficiency of the joint estimator for location and shape}

As previously anticipated, this last sub session is devoted to the assessment of the semiparametric efficiency of the joint estimator $(\hat{\bs{\mu}}_{Ty},\widehat{\mb{V}}_{1,R})$, where $\hat{\bs{\mu}}_{Ty}$ is the Tyler's estimator in Eq. \eqref{Ty_mu} of the location vector $\bs{\mu}_0$, while $\widehat{\mb{V}}_{1,R}$ is the $R$-estimator in Eq. \eqref{mat_one_step_R_new_Ty} of the shape matrix $\mb{V}_{1,0}$ exploiting the Tyler's joint estimator $(\hat{\bs{\mu}}_{Ty},\widehat{\mb{V}}_{1,Ty})$ as preliminary $\sqrt{L}$-consistent estimators. 

As basis of comparison, we also report the performance of the joint \virg{sample} estimator $(\hat{\bs{\mu}}_{SM},\widehat{\mb{V}}_{1,SCM})$, defined as:
\be
\label{SM_mu}
\hat{\bs{\mu}}_{SM} \triangleq L^{-1} \sum\nolimits_{l=1}^{L}\mb{z}_l,
\ee
\be
\label{SCM_V}
\left\lbrace  \begin{array}{l}
	\widehat{\bs{\Sigma}}_{SCM} \triangleq L^{-1} \sum\nolimits_{l=1}^{L}(\mb{z}_l-\hat{\bs{\mu}}_{SM})(\mb{z}_l-\hat{\bs{\mu}}_{SM})^H\\
	\widehat{\mb{V}}_{1,SCM} \triangleq \nicefrac{\widehat{\bs{\Sigma}}_{SCM}}{[\widehat{\bs{\Sigma}}_{SCM}]_{1,1}}.
\end{array}\right. 
\ee

The performance assessment will be performed in terms of the following indices:\\
\textit{Bias indices}
\begin{itemize}
	\item Bias index for the estimation of the location vector $\bs{\mu}_0$:
	\be
	\beta_\gamma \triangleq \norm{E\{\hat{\bs{\mu}}_\gamma-\bs{\mu}_0\}}_2,
	\ee
	where $\gamma \in \{SM,\; Ty\}$ indicates the sample mean in Eq. \eqref{SM_mu} or the Tyler's estimator in Eq. \eqref{Ty_mu}.
	\item Bias index for the estimation of the shape matrix $\mb{V}_{1,0}$:
	\be
	\varphi_\gamma \triangleq \norm{E\{\ovec{\widehat{\mb{V}}_{1,\gamma}-\mb{V}_{1,0}}\}}_2,
	\ee
	where $\gamma \in \{SCM,\; Ty, R-vdW, R-t_5\}$ indicates a specif estimator among the SCM in Eq. \eqref{SCM_V}, the Tyler's shape estimator in Eq. \eqref{Ty_schape} and the $R$-estimator in Eq. \eqref{mat_one_step_R_new_Ty} that relies on the Tyler's one as preliminary estimator. Note that for the $R$-estimator we have two options: $\widehat{\mb{V}}_{1,{R-vdW}}$ indicates the $R$-estimator in Eq. \eqref{mat_one_step_R_new_Ty} exploiting the \textit{van der Waerden} score in Eq. \eqref{K_van} while $\widehat{\mb{V}}_{1,{R-t_5}}$ indicates again $R$-estimator in Eq. \eqref{mat_one_step_R_new_Ty} but exploiting the $t$-score in Eq. \eqref{K_tnu} with $\nu =5$.
\end{itemize}

\textit{Mean Squared Error (MSE) indices}
\begin{itemize}
	\item MSE index for the estimation of the location vector $\bs{\mu}_0$:
	\be
	\varrho_\gamma \triangleq \norm{E\{(\hat{\bs{\mu}}^a_\gamma-\bs{\mu}^a_0)(\hat{\bs{\mu}}^a_\gamma-\bs{\mu}^a_0)^H\}}_F,
	\ee
	where $\gamma \in \{SM,\; Ty\}$ and for a given $\mb{x}\in \mathbb{C}^N$, $\mb{x}^a\triangleq (\mb{x}^T,\mb{x}^H)^T \in \mathbb{C}^{2N}$.
	\item MSE index for the estimation of the shape matrix $\mb{V}_{1,0}$:
	\be
	\varsigma_\gamma \triangleq \norm{E\{\ovec{	\widehat{\mb{V}}_{1,\gamma}-\mb{V}_{1,0}}\ovec{	\widehat{\mb{V}}_{1,\gamma}-\mb{V}_{1,0}}^H\}}_F,
	\ee
	and $\gamma \in \{SCM,\; Ty, R-vdW, R-t_5\}$, as before, indicates the relevant estimator at hand.
\end{itemize}

As lower bound indices, we use
\be
\varepsilon_{SCRB,\bs{\mu}_0} \triangleq \norm{\mathrm{SCRB}(\bs{\mu}_0|h_0)}_F,
\ee
\be
\varepsilon_{SCRB,\mb{V}_{1,0}} \triangleq \norm{\mathrm{SCRB}({\ovec{\mb{V}_{1,0}}}|h_0)}_F,
\ee
where $\mathrm{SCRB}(\bs{\mu}_0|h_0)$ is given in \eqref{SCRB_mu} and $\mathrm{SCRB}({\ovec{\mb{V}_{1,0}}}|h_0)$ in \eqref{SCRB_shape}.

The bias indices of the sample mean in Eq. \eqref{SM_mu} and of the Tyler's estimator in Eq. \eqref{Ty_schape} are reported in Fig. \ref{fig:Fig1}. As we can note, the bias is on the order of $10^{-3}$, so it can be considered negligible and the two estimators unbiased. Fig. \ref{fig:Fig3} shows the MSE performance of the sample mean $\hat{\bs{\mu}}_{SM}$ estimator in \eqref{SM_mu} and of the Tyler's estimator $\hat{\bs{\mu}}_{Ty}$ in \eqref{Ty_mu} compared to the lover bound in \eqref{SCRB_mu}. As wee can see, $\hat{\bs{\mu}}_{Ty}$ is almost efficient with respect to $\mathrm{SCRB}(\bs{\mu}_0|h_0)$ in heavy-tailed data ($0<s<1$) and outperforms $\hat{\bs{\mu}}_{SM}$ that it is known to be non robust. On the other hand, $\hat{\bs{\mu}}_{SM}$ is efficient in the Gaussian case ($s=1$), and tends to have better performance than $\hat{\bs{\mu}}_{Ty}$ for $s>1$. However, in this light-tails scenario, the MSE of $\hat{\bs{\mu}}_{Ty}$ does not explode and remains close to the $\hat{\bs{\mu}}_{SM}$'s one. 

\begin{figure}[h]
	\centering
	\includegraphics[height=5cm]{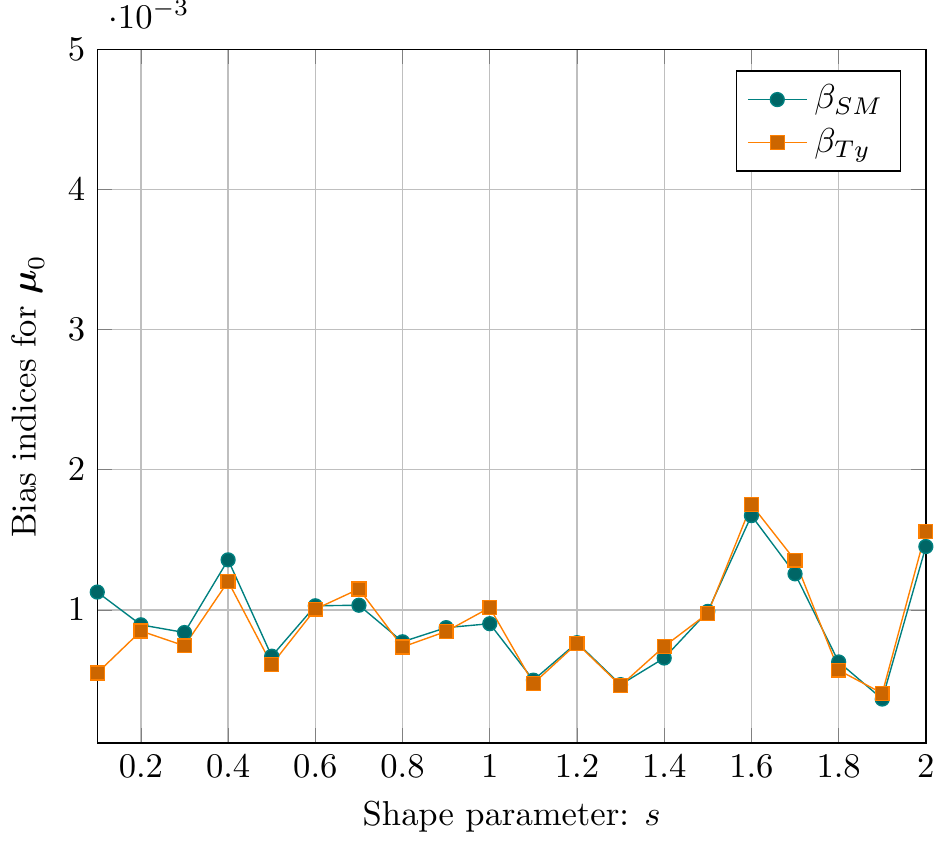}
	\caption{Bias in the estimation of $\bs{\mu}_0$ for the sample mean and for the Tyler's estimator.}
	\label{fig:Fig2}
\end{figure}

\begin{figure}[h]
	\centering
	\includegraphics[height=5cm]{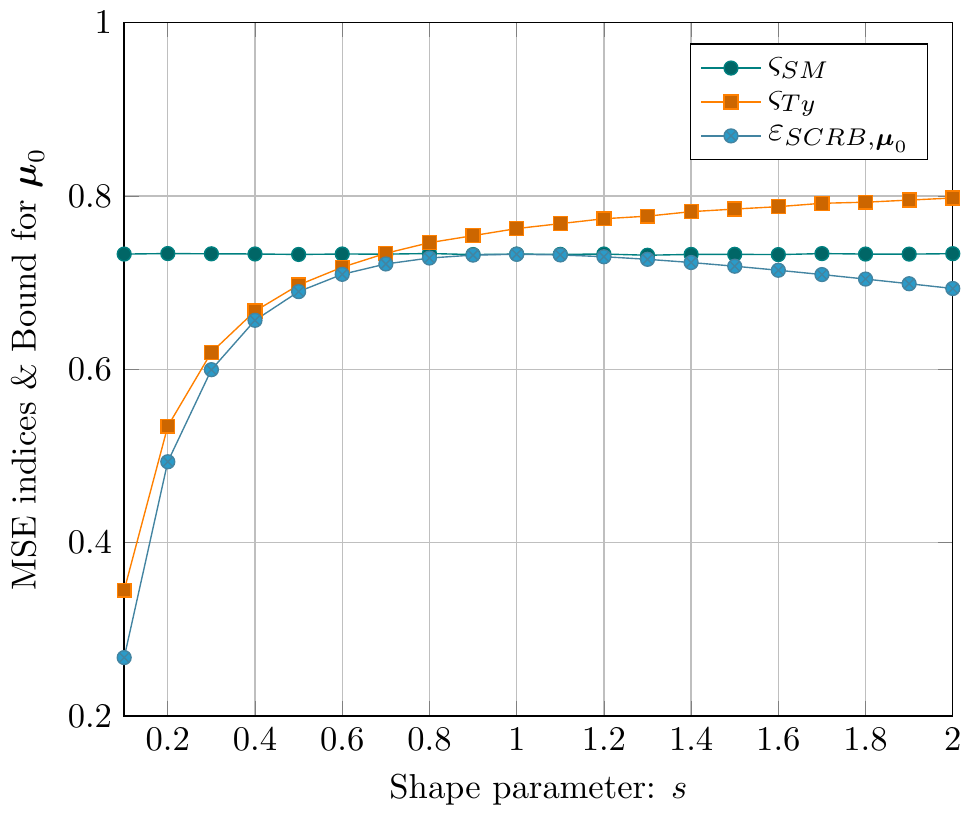}
	\caption{MSE in the estimation of $\bs{\mu}_0$ for the sample mean and for the Tyler's estimator.}
	\label{fig:Fig3}
\end{figure}

\begin{figure}[h]
	\centering
	\includegraphics[height=5cm]{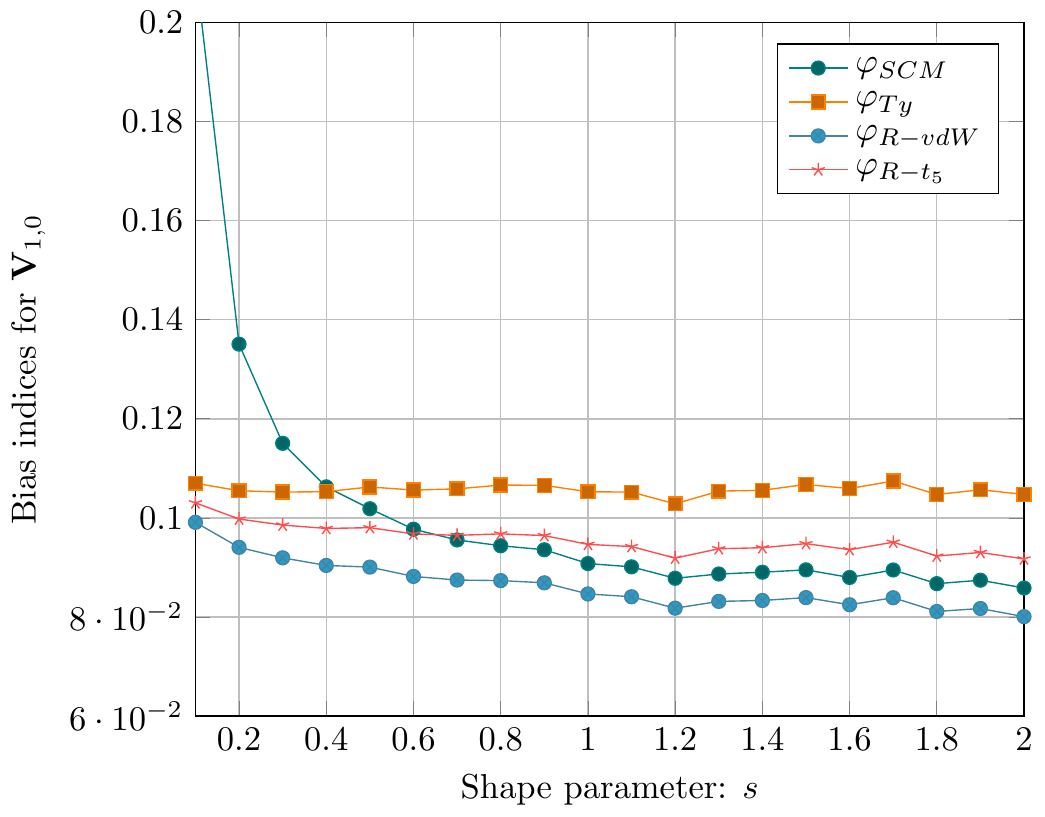}
	\caption{Bias in the estimation of $\mb{V}_{1,0}$ for the SCM, the Tyler's estimator and the $R$-estimator exploiting both the van der Waerden and the $t_5$-score functions.}
	\label{fig:Fig4}
\end{figure}

\begin{figure}[h]
	\centering
	\includegraphics[height=5cm]{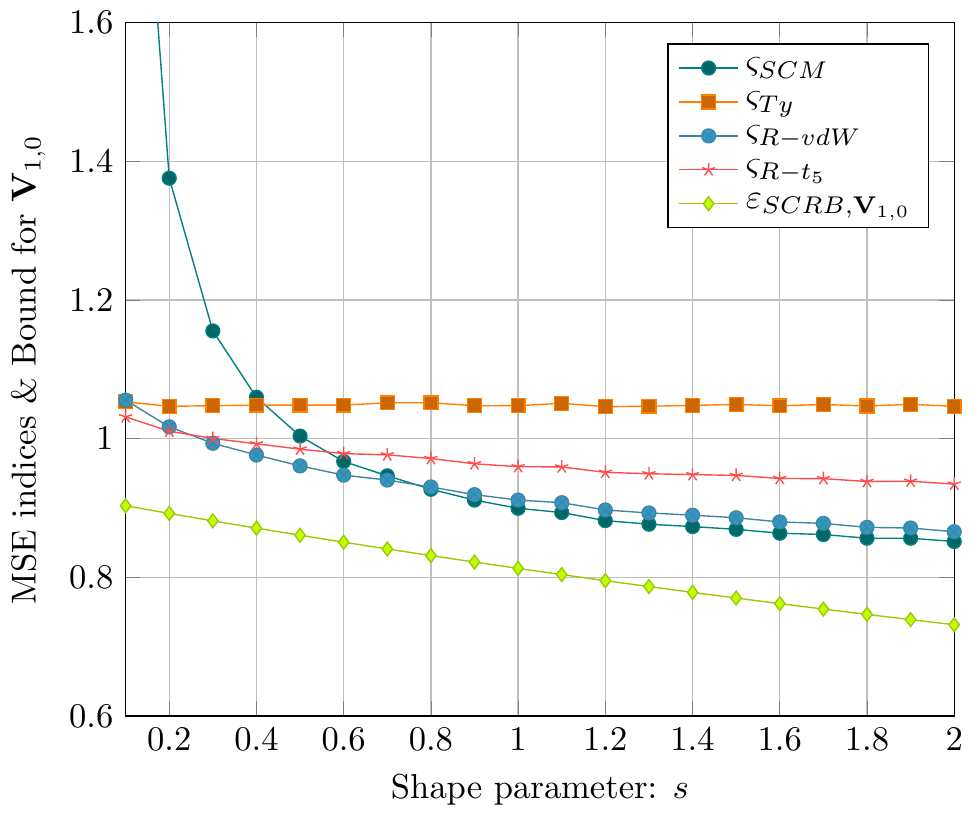}
	\caption{MSE in the estimation of $\mb{V}_{1,0}$ for the SCM, the Tyler's estimator and the $R$-estimator exploiting both the van der Waerden and the $t_5$-score functions..}
	\label{fig:Fig5}
\end{figure}

As far it concern the shape matrix estimation, the simulation results are shown in Fig. \ref{fig:Fig4} for the bias and Fig. \ref{fig:Fig5} for the MSE. The main fact here is that the two $R$-estimators $\widehat{\mb{V}}_{1,{R-vdW}}$ and $\widehat{\mb{V}}_{1,{R-t_5}}$ in Eq. \eqref{mat_one_step_R_new_Ty} outperforms the Tyler's estimator $\widehat{\mb{V}}_{1,Ty}$ in \eqref{Ty_schape} for every values of $s$, i.e. for both heavy-tailed and light-tailed data. Moreover, as expected, $\widehat{\mb{V}}_{1,{R-vdW}}$ and $\widehat{\mb{V}}_{1,{R-t_5}}$ greatly outperform the sample covariance matrix $\widehat{\mb{V}}_{1,SCM}$ in Eq. \eqref{SCM_V} in the presence of heavy-tailed data ($0<s<1$), while their MSE is of the same order for $s>1$. Between the two $R$-estimators, we can notice that $\widehat{\mb{V}}_{1,{R-vdW}}$ has better performance than $\widehat{\mb{V}}_{1,{R-t_5}}$ in terms of both bias and MSE. Finally, a comment on the efficiency of the above-mentioned estimator is in order. As we can see, there is a gap between the MSE indices of $\widehat{\mb{V}}_{1,SCM}$, $\widehat{\mb{V}}_{1,Ty}$ and $\widehat{\mb{V}}_{1,{R-vdW}}$ and $\widehat{\mb{V}}_{1,{R-t_5}}$ and the SCRB. However, it is worth to underline that our aim here is to compare the performance of shape matrix estimators in a \virg{finite-sample} regime, i.e. with a number of observations equal to $L=5N$ that represents a reasonable value in many practical applications. Of course, by letting $L \rightarrow \infty$, it can be shown that both the two $R$-estimators $\widehat{\mb{V}}_{1,{R-vdW}}$ and $\widehat{\mb{V}}_{1,{R-t_5}}$ achieve the bound $\mathrm{SCRB}({\ovec{\mb{V}_{1,0}}}|h_0)$ in Eq. \eqref{SCRB_shape} as predicted by theoretical considerations \cite{Hallin_Annals_Stat_2,Sem_eff_est_TSP}.

In summary, previous simulations highlights the benefits that the proposed robust $R$-estimator can bring. Specifically, it always outperforms the Tyler's estimator in both heavy- and light-tails scenarios. Moreover its estimation performance is way better that the SCM one in heavy-tailed data while it is almost similar in light-tailed scenarios: high gain, very small loss. These very promising results promote the use of the $R$-estimator to other problems, as the structured shape estimation discussed in \cite{MERIAUX,Meriaux_TSP}.

\section{Conclusions}
\label{conclusions}
This paper dealt with the fundamental problem of estimating the location vector and the shape matrix of a set of CES distributed data. In the first part of this work, we derived a computationally efficient version of the robust and semiparametric efficient $R$-estimator already proposed in \cite{Sem_eff_est_TSP}. Remarkably, the new \virg{matrix version} of the $R$-estimator can provide the same estimate of the shape matrix but at a computational time that is more than two order of magnitude smaller with respect to the \virg{vectorized version} previously derived in \cite{Sem_eff_est_TSP}. This fundamental property suggests us to use the new version given in Eq. \eqref{mat_one_step_R_new_1} as the default version of the $R$-estimator. In the second part of this paper, the joint estimation of the location vector $\bs{\mu}_0$ and the shape matrix $\mb{V}_{1,0}$ of a set of i.i.d.\ CES-distributed, multivariate observations has been addressed. Building upon the asymptotic decorrelation of the location and shape estimation problems, a joint estimator that relies on the Tyler's $M$-estimator $\hat{\bs{\mu}}_{Ty}$ for $\bs{\mu}_0$ and on a recently proposed $R$-estimator $\widehat{\mb{V}}_{1,R}$ for $\mb{V}_{1,0}$ has been discussed and its performance, in terms of both bias and MSE, assessed and compared with the relevant Semiparametric Cram\'er-Rao Bound. Our simulation results, obtained for GG-distributed data, have shown that joint estimator $(\hat{\bs{\mu}}_{Ty},\widehat{\mb{V}}_{1,R})$ of location and shape represents a good alternative to the classical Maronna's joint $M$-estimators. In particular, in terms of shape matrix estimation, the proposed joint estimator outperforms the joint Tyler's estimator in both heavy-tailed and light-tailed data. Future works will investigate the application of the proposed estimator in robust clustering and distance learning problems.

\appendix\normalsize
\setcounter{equation}{0}
\renewcommand\theequation{A.\arabic{equation}}
\section{Appendix: Proof of the Eqs. \eqref{mat_one_step_R_new_1} and \eqref{com_alpha_hat_comp_1}}

The main aim of this Appendix is to show how to obtain a closed from expression of the (complex-valued) $R$-estimator for the shape matrix $\mb{V}_1$, provided in \eqref{R_est} (see also \cite[Eq. (54)]{Sem_eff_est_TSP}) by avoiding the use of the matrix $\mb{L}_{\mb{V}_1}$. The matrix $\mb{L}_{\mb{V}_1}$ is in fact a structured $(N^2-1) \times (N^2-1)$ that is built upon the $N \times N$ matrix $\mb{V}_1$. Consequently, if we are able to obtain a expression of the $R$-estimator that relies only on $\mb{V}_1$ and not on $\mb{L}_{\mb{V}_1}$, we would gain a lot in terms of computational efficiency.

At first, let us recall here the expression of the $R$-estimator introduced in Eqs. \eqref{R_est} and \eqref{com_alpha_hat_1}:
\begin{equation*}
\begin{split}
		\ovec{\widehat{\mb{V}}_{1,R}} & = \ovec{\widehat{\mb{V}}_1^\star} + \\
		&\frac{1}{L\hat{\alpha}_\mathbb{C}}\quadre{\mb{L}_{\widehat{\mb{V}}_1^\star} \mb{L}_{\widehat{\mb{V}}_1^\star}^\mathsf{H}}^{-1}
		\mb{L}_{\widehat{\mb{V}}_1^\star}\sum\nolimits_{l=1}^{L}K_h\tonde{\frac{r_l^\star}{L+1}} \mathrm{vec}(\hat{\mb{u}}^\star_l(\hat{\mb{u}}^\star_l)^\mathsf{H}).
\end{split}
\end{equation*}
where
\begin{equation*}
\hat{\alpha}_\mathbb{C} = \nicefrac{\norm{\widetilde{\bs{\Delta}}^\mathbb{C}_{\widehat{\mb{V}}_1^\star + L^{-1/2}\mb{H}^0_\mathbb{C}} - \widetilde{\bs{\Delta}}^\mathbb{C}_{\widehat{\mb{V}}_1^\star}}}{ \norm{ \mb{L}_{\widehat{\mb{V}}_1^\star} \mb{L}_{\widehat{\mb{V}}_1^\star}^\mathsf{H}\ovec{\mb{H}^0_\mathbb{C}}} },
\end{equation*}
and $\mb{H}^0_\mathbb{C}$ is a \virg{small perturbation}, Hermitian, matrix s. t. $[\mb{H}^0_\mathbb{C}]_{1,1}=0$.

In the calculation proposed below, we make extensive use of the following properties holding for conforming matrices (see e.g. \cite{Matrix_Cook}):
\be\label{p_vec_tr}
\trace{\mb{A}\mb{B}} = \vc{\mb{A}^\top}^\top\vc{\mb{B}}
\ee
\be\label{p_kr_vec}
\vc{\mb{A}\mb{X}\mb{B}} = (\mb{B}^\top \otimes \mb{A})\vc{\mb{X}}
\ee
\be
(\mb{A} \otimes \mb{B})(\mb{C} \otimes \mb{D}) = (\mb{AC} \otimes \mb{BD})
\ee
\be
(\mb{A} \otimes \mb{B})^\top = \mb{A}^\top \otimes \mb{B}^\top, \quad (\mb{A} \otimes \mb{B})^\topH = \mb{A}^\topH \otimes \mb{B}^\topH
\ee
 
\subsection{Matrix version of the central sequence}
\label{sec:_cent_seq_mv}

The \virg{distribution-free} version of the (complex-valued) efficient central sequence $\widetilde{\bs{\Delta}}_{\widehat{\mb{V}}_1^\star}^{\mathbb{C}}$ is defined in Eq. \eqref{complex_app_eff_cs_1} as: 

\begin{equation*}
\widetilde{\bs{\Delta}}_{\widehat{\mb{V}}_1^\star}^{\mathbb{C}} \triangleq \frac{1}{\sqrt{L}}\mb{L}_{\widehat{\mb{V}}_1^\star}\sum_{l=1}^{L}K_h\tonde{\frac{r_l^\star}{L+1}}  \mathrm{vec}(\hat{\mb{u}}^\star_l(\hat{\mb{u}}^\star_l)^\topH).
\end{equation*}

Let us start by showing how to re-write $\widetilde{\bs{\Delta}}_{\widehat{\mb{V}}_1^\star}^{\mathbb{C}}$ without using $\mb{L}_{\mb{V}_1}$.  

For notation simplicity, we introduce the scalar $\kappa_l$ as:
\be\label{k_reduced}
\kappa_l \triangleq \frac{1}{\sqrt{L}} K_h\tonde{\frac{r_l^\star}{L+1}}
\ee

Let us now define the $N^2$-dimensianl vector $\mb{z}_{\widehat{\mb{V}}_1^\star}$ as:
\be\label{evv_cs_1}
\mb{z}_{\widehat{\mb{V}}_1^\star} = \quadre{z_1,\quad (\underline{\mb{z}}_{\widehat{\mb{V}}_1^\star})^\top}^\top = \quadre{z_1,\quad (\widetilde{\bs{\Delta}}_{\widehat{\mb{V}}_1^\star}^{\mathbb{C}})^\top}^\top.
\ee
where $z_1$ is an unspecified complex number. Then, from Eq. \eqref{complex_app_eff_cs_1} and from the definition of the matrices $\mb{P}$ and $\mb{L}_{\mb{V}_1}$ in Eqs. \eqref{mat_P} and \eqref{L_mat} respectively, the vector $\mb{z}_{\widehat{\mb{V}}_1^\star}$ can be cast as:
\be\label{mod_z}
\begin{split}
	\mb{z}_{\widehat{\mb{V}}_1^\star} 
	&= \quadre{\kronVtinvmTS} \Pi^{\perp}_{\cvec{\mb{I}_N}} \sum_{l=1}^{L}\kappa_l\mathrm{vec}(\hat{\mb{u}}^\star_l(\hat{\mb{u}}^\star_l)^\topH)\\
	&= \quadre{\kronVtinvmTS} \quadre{\sum_{l=1}^{L}\kappa_l\mathrm{vec}(\hat{\mb{u}}^\star_l(\hat{\mb{u}}^\star_l)^\topH) - \frac{1}{N}\vc{\mb{I}_N}\sum_{l=1}^{L}\kappa_l}\\
	& = \mathrm{vec}\tonde{(\pV)^{-1/2} \mb{R} (\pV)^{-1/2} - \zeta (\pV)^{-1}}
\end{split}
\ee
where:
\be\label{def_R}
\mb{R} \triangleq \sum_{l=1}^{L}\kappa_l\hat{\mb{u}}^\star_l(\hat{\mb{u}}^\star_l)^\topH,
\ee
and
\be\label{def_zeta} 
\zeta \triangleq  \frac{1}{N}\sum_{l=1}^{L}\kappa_l.
\ee
Consequently, we have that:
\be
\label{z_vec}
\widetilde{\bs{\Delta}}_{\widehat{\mb{V}}_1^\star}^{\mathbb{C}}  =  \underline{\mathrm{vec}}\tonde{(\pV)^{-1/2} \mb{R} (\pV)^{-1/2} - \zeta (\pV)^{-1}} \triangleq \underline{\mb{z}}_{\widehat{\mb{V}}_1^\star}.
\ee

Note that in \eqref{mod_z}, we have used the fact that:
\be
\trace{(\hat{\mb{u}}^\star_l(\hat{\mb{u}}^\star_l)^\topH} = (\hat{\mb{u}}^\star_l)^\topH\hat{\mb{u}}^\star_l = 1, \quad \forall l.
\ee

To avoid confusion, we indicate as $\underline{\mb{z}}_{\widehat{\mb{V}}_1^\star}$ the \virg{$\mb{L}_{\widehat{\mb{V}}_1^\star}$-free} version of $\widetilde{\bs{\Delta}}_{\widehat{\mb{V}}_1^\star}^{\mathbb{C}}$ obtained in Eq. \eqref{z_vec}.
It is important to underline in fact that $\underline{\mb{z}}_{\widehat{\mb{V}}_1^\star}$ does not make use of the \virg{unnecessary large} $(N^2-1) \times (N^2-1)$ matrix $\mb{L}_{\widehat{\mb{V}}_1^\star}$, while only $N \times N$ matrices are involved.

\subsection{An \virg{$\mb{L}_{\widehat{\mb{V}}_1^\star}$-free} version of the scalar $\hat{\alpha}_\mathbb{C}$}

Let us define the $N^2$-dimensional vector as:
\be\label{r_1}
\mb{r}_{\widehat{\mb{V}}_1^\star} = \quadre{r_1,\quad (\underline{\mb{r}}_{\widehat{\mb{V}}_1^\star})^\top}^\top = \quadre{r_1,\quad (\mb{L}_{\widehat{\mb{V}}_1^\star} \mb{L}_{\widehat{\mb{V}}_1^\star}^\mathsf{H}\ovec{\mb{H}^0_\mathbb{C}})^\top}^\top,
\ee
where $r_1$ is an unspecified complex scalar.

By using the fact that, by definition, $[\mb{H}^0_\mathbb{C}]_{1,1}=0$, we have that:
\be
\begin{split}
	\mb{r}_{\widehat{\mb{V}}_1^\star}^0 &= \quadre{r_1^0, \quad (\mb{L}_{\widehat{\mb{V}}_1^\star} \mb{L}_{\widehat{\mb{V}}_1^\star}^\mathsf{H}\ovec{\mb{H}^0_\mathbb{C}})^\top}^\top \\
	& = \quadre{\tonde{\kronVtinvmTS}\Pi^{\perp}_{\cvec{\mb{I}_N}}\tonde{\kronVtinvmTS}^\topH} \vc{\mb{H}^0_\mathbb{C}}\\
	& = \quadre{\kronVtinvTS - N^{-1} \mathrm{vec}\tonde{(\pV)^{-1}}\mathrm{vec}\tonde{(\pV)^{-1}}^\topH}\vc{\mb{H}^0_\mathbb{C}}\\
	& = \mathrm{vec}\tonde{(\pV)^{-1} \mb{H}^0_\mathbb{C} (\pV)^{-1} -N^{-1} \mathrm{tr}\tonde{(\pV)^{-1} \mb{H}^0_\mathbb{C}} (\pV)^{-1}}
\end{split}
\ee

Then, from Eq. \eqref{r_1}, we get:
\be
\underline{\mb{r}}_{\widehat{\mb{V}}_1^\star}^0 = \underline{\mathrm{vec}}\tonde{(\pV)^{-1} \mb{H}^0_\mathbb{C} (\pV)^{-1} -N^{-1} \mathrm{tr}\tonde{(\pV)^{-1} \mb{H}^0_\mathbb{C}} (\pV)^{-1}},
\ee
that depends only on $N \times N$ matrix quantities.

Finally, by using the previous results, an \virg{$\mb{L}_{\widehat{\mb{V}}_1^\star}$-free} version of $\hat{\alpha}_\mathbb{C}$ can be expressed as:

\be\label{com_alpha_hat_comp}
	\hat{\alpha}_\mathbb{C} = \frac{\norm{\widetilde{\bs{\Delta}}^\mathbb{C}_{\widehat{\mb{V}}_1^\star + L^{-1/2}\mb{H}^0_\mathbb{C}} - \widetilde{\bs{\Delta}}^\mathbb{C}_{\widehat{\mb{V}}_1^\star}}}{ \norm{ \mb{L}_{\widehat{\mb{V}}_1^\star} \mb{L}_{\widehat{\mb{V}}_1^\star}^\mathsf{H}\ovec{\mb{H}^0_\mathbb{C}}} }
	 = \frac{\norm{\underline{\mb{z}}_{\widehat{\mb{V}}_1^\star + L^{-1/2}\mb{H}^0_\mathbb{C}} - \underline{\mb{z}}_{\widehat{\mb{V}}_1^\star}}}{\norm{\underline{\mb{r}}_{\widehat{\mb{V}}_1^\star}^0}}.
\ee

\subsection{An \virg{$\mb{L}_{\widehat{\mb{V}}_1^\star}$-free} matrix version of the $R$-estimator}

By using the scalar $\kappa_l$, previously defined in Eq. \eqref{k_reduced}, we can re-write the expression of the $R$-estimator in Eq. \eqref{R_est} as:
\be
\label{ovec_one_step_R}
\begin{split}
	\ovec{\widehat{\mb{V}}_{1,R}} & = \ovec{\widehat{\mb{V}}_1^\star} +\frac{1}{\sqrt{L}\hat{\alpha}_\mathbb{C}}\quadre{\mb{L}_{\widehat{\mb{V}}_1^\star} \mb{L}_{\widehat{\mb{V}}_1^\star}^\mathsf{H}}^{-1} \mb{L}_{\widehat{\mb{V}}_1^\star}  \sum_{l=1}^{L} \kappa_l \mathrm{vec}(\hat{\mb{u}}^\star_l(\hat{\mb{u}}^\star_l)^\mathsf{H}).
\end{split}
\ee

By using the \virg{$\mb{L}_{\widehat{\mb{V}}_1^\star}$-free} version of $\widetilde{\bs{\Delta}}_{\widehat{\mb{V}}_1^\star}^{\mathbb{C}}$ obtained in Eq. \eqref{z_vec}, the expression in Eq. \eqref{ovec_one_step_R} can be rewritten as:
\be
\label{vec_one_step_R}
\begin{split}
	\vc{\widehat{\mb{V}}_{1,R}} & = \vc{\widehat{\mb{V}}_1^\star} +\frac{1}{\sqrt{L}\hat{\alpha}_\mathbb{C}} \mb{P}^\top \quadre{\mb{L}_{\widehat{\mb{V}}_1^\star}\mb{L}_{\widehat{\mb{V}}_1^\star}^\mathsf{H}}^{-1}\mb{P} \times \\
	& \mathrm{vec}\tonde{(\pV)^{-1/2} \mb{R}(\pV)^{-1/2} - \zeta (\pV)^{-1}}.
\end{split}
\ee

To get rid of the \virg{unnecessary large} $N^2 \times N^2$ matrix $\mb{P}^\top \quadre{\mb{L}_{\widehat{\mb{V}}_1^\star}\mb{L}_{\widehat{\mb{V}}_1^\star}^\mathsf{H}}^{-1}\mb{P}$ we make use of the extension to the complex field of the result obtained in Lemma 3.1 of \cite{Hallin_P_Annals}.  
%
Specifically, it can be shown that (see \cite[Appendix A.2]{Hallin_Annals_Stat_2}):
\be\label{lemma_hallin_ex}
\begin{split}
	\mb{P}^\top &\quadre{\mb{L}_{\widehat{\mb{V}}_1^\star} \mb{L}_{\widehat{\mb{V}}_1^\star}^\mathsf{H}}^{-1} \mb{P}\\
	&= \quadre{\mb{I}_{N^2}-\vc{\widehat{\mb{V}}_1^\star}\mb{e}_1^\top}\quadre{\kronVTS}\quadre{\mb{I}_{N^2}-\vc{\widehat{\mb{V}}_1^\star}\mb{e}_1^\top}^\topH,
\end{split}
\ee
where $\mb{e}_1$ is the first vector of the canonical basis of $\mathbb{R}^{N^2}$.

Through direct calculation, we can easily show that:
\be
\begin{split}
	\mb{P}^\top &\quadre{\mb{L}_{\widehat{\mb{V}}_1^\star} \mb{L}_{\widehat{\mb{V}}_1^\star}^\mathsf{H}}^{-1} \mb{P} \mathrm{vec}\tonde{(\pV)^{-1/2}
		 \mb{R}(\pV)^{-1/2} - \zeta (\pV)^{-1}}\\
	 &= \quadre{\mb{I}_{N^2}-\vc{\widehat{\mb{V}}_1^\star}\mb{e}_1^\top}\quadre{\kronVTS} \times \\
	 &\qquad\qquad \mathrm{vec}\tonde{(\pV)^{-1/2} \mb{R}(\pV)^{-1/2} - \zeta (\pV)^{-1}}\\
	 &= \quadre{\mb{I}_{N^2}-\vc{\widehat{\mb{V}}_1^\star}\mb{e}_1^\top}\mathrm{vec}\tonde{(\pV)^{1/2} \mb{R}(\pV)^{1/2} - \zeta \pV}\\
	 &= \mathrm{vec}\tonde{\mb{W} - \quadre{\mb{W}}_{1,1}\widehat{\mb{V}}_1^\star},
\end{split}
\ee
where
\be\label{mat_W}
\begin{split}
	\mb{W} & \triangleq L^{1/2}(\pV)^{1/2} \mb{R}(\pV)^{1/2} \\
	& = (\pV)^{1/2} \quadre{\frac{1}{L}\sum_{l=1}^{L}K_h\tonde{\frac{r_l^\star}{L+1}}\hat{\mb{u}}^\star_l(\hat{\mb{u}}^\star_l)^\topH} (\pV)^{1/2}.
\end{split}
\ee

By collecting the previous results, the $R$-estimator in Eq. \eqref{R_est} can be expressed as:
\be
\label{vec_one_step_R_new}
\begin{split}
	\vc{\widehat{\mb{V}}_{1,R}} & = \vc{\widehat{\mb{V}}_1^\star} +\frac{1}{\hat{\alpha}_\mathbb{C}} \vc{\mb{W} - \quadre{\mb{W}}_{1,1}\widehat{\mb{V}}_1^\star}.
\end{split}
\ee
Finally, in matrix form, we have:
\be
\label{mat_one_step_R_new}
\boxed{\widehat{\mb{V}}_{1,R} = \widehat{\mb{V}}_1^\star +\frac{1}{\hat{\alpha}_\mathbb{C}} \tonde{\mb{W} - \quadre{\mb{W}}_{1,1}\widehat{\mb{V}}_1^\star}},
\ee
where
\be
\boxed{\hat{\alpha}_\mathbb{C} = \frac{\norm{\underline{\mb{z}}_{\widehat{\mb{V}}_1^\star + L^{-1/2}\mb{H}^0_\mathbb{C}} - \underline{\mb{z}}_{\widehat{\mb{V}}_1^\star}}}{\norm{\underline{\mb{r}}_{\widehat{\mb{V}}_1^\star}^0}}}.
\ee
This concludes the proof of Eqs. \eqref{mat_one_step_R_new_1} and \eqref{com_alpha_hat_comp_1}.

%
%


%
%

\bibliographystyle{spbasic}      
\bibliography{ref_semipar_eff_estim}   

\end{document}